\documentclass[fleqn,usenatbib]{mnras}

\newcommand{\refereenew}{}
\newcommand{\january}{}

\usepackage[T1]{fontenc}
\usepackage{ae,aecompl}

\usepackage{graphicx}	% Including figure files
\usepackage{amsmath}	% Advanced maths commands
\usepackage{amssymb}	% Extra maths symbols
\usepackage{amsbsy}
\usepackage{mathtools} % bmatrix*
\usepackage{url}
\usepackage[usenames,dvipsnames]{color}
\usepackage{xspace}
\usepackage{siunitx}
\DeclareSIUnit\percentage{\,per\>cent}  % Spell out percent as per cent.
\DeclareSIUnit\erg{erg}  % Declare the unit erg
\DeclareSIUnit\solarradius{\ensuremath{\text{R}_\odot}}
\DeclareSIUnit\solarmass{\ensuremath{\text{M}_\odot}}
\DeclareSIUnit\solarlum{\ensuremath{\text{L}_\odot}}
\DeclareSIUnit\angstrom{\ensuremath{\textup{\AA}}}
\DeclareSIUnit\parsec{\ensuremath{\text{pc}}}

\usepackage{newtxtext,newtxmath}
\newcommand{\figref}[1]{Fig.~\ref{#1}}
\newcommand{\secref}[1]{Sec.~\ref{#1}}
\newcommand{\flag}[1]{\texttt{#1}}

\newcommand{\numax}{\ensuremath{\nu_\text{max}}\xspace}
\newcommand{\dnu}{\ensuremath{\Delta\nu}\xspace}

\newcommand{\teff}{\ensuremath{T_{\textup{eff}}}\xspace}
\newcommand{\feh}{\ensuremath{[\text{Fe/H}]\xspace}}
\newcommand{\meh}{\ensuremath{[\text{M/H}]\xspace}}

\newcommand{\alphafe}{\ensuremath{[\text{$\alpha$/Fe}]}\xspace}
\newcommand{\parallax}{\ensuremath{\varpi}\xspace}
\newcommand{\Jr}{\ensuremath{J_{R}}\xspace}
\newcommand{\Jz}{\ensuremath{J_{z}}\xspace}
\newcommand{\Jphi}{\ensuremath{J_{\phi}}\xspace}
\newcommand{\Lz}{\ensuremath{L_{z}}\xspace}
\newcommand{\thetar}{\ensuremath{\theta_{r}}\xspace}
\newcommand{\thetaz}{\ensuremath{\theta_{z}}\xspace}
\newcommand{\thetaphi}{\ensuremath{\theta_{\phi}}\xspace}

\newcommand{\data}{\ensuremath{\mathbf{x}}\xspace}
\newcommand{\error}{\ensuremath{\mathbf{\sigma}}\xspace}
\newcommand{\hyper}{\ensuremath{\lambda}\xspace}
\newcommand{\nc}{\ensuremath{n_{c}}\xspace}
\newcommand{\gweight}{\ensuremath{w_i}\xspace}
\newcommand{\gmean}{\ensuremath{\mu_i}\xspace}
\newcommand{\gcov}{\ensuremath{\mathbf{\Sigma}_i}\xspace}

\newcommand{\galr}{\ensuremath{R_{\textup{GC}}}\xspace}

\newcommand{\kepler}{\emph{Kepler}\xspace}
\newcommand{\gaia}{\emph{Gaia}\xspace}
\newcommand{\basta}{\textsc{basta}\xspace}
\newcommand{\corot}{{\sc CoRoT}\xspace}
\newcommand{\ktwo}{{\it K2}\xspace}
\newcommand{\tess}{{\sc TESS}\xspace}
\newcommand{\syd}{\textsc{SYD}\xspace}

\newcommand{\lamost}{LAMOST\xspace}
\newcommand{\apogee}{APOGEE\xspace}
\newcommand{\galah}{GALAH\xspace}
\newcommand{\tmass}{2MASS\xspace}

\newcommand{\ktwocn}{1-8, 10-18\xspace}

\newcommand{\noofstars}{21,076\xspace}
\newcommand{\noofstarsbeforepruning}{22,711\xspace}
\newcommand{\noofremovedbasta}{1596\xspace}
\newcommand{\noofkeplerstars}{12424\xspace}
\newcommand{\noofktwostars}{8220\xspace}
\newcommand{\noofcorotstars}{432\xspace}
\newcommand{\noofapogeestars}{10,071\xspace}
\newcommand{\nooflamoststars}{7164\xspace}
\newcommand{\noofgalahstars}{3841\xspace}

\newcommand{\ruwepercent}{92.57\xspace}
\newcommand{\ruwepercentsingle}{86.93\xspace}
\newcommand{\knownphase}{20,129\xspace}

\newcommand{\youngthickred}{red\xspace}
\newcommand{\thickorange}{orange\xspace}
\newcommand{\youngthingreen}{green\xspace}
\newcommand{\thinpurple}{purple\xspace}
\newcommand{\haloblue}{blue\xspace}
\newcommand{\oldthickolive}{olive\xspace}

\title[Chronology of the Galactic disc]{A Unified Exploration of the Chronology of the Galaxy}

\newcommand{\affdifa}{1}
\newcommand{\affinaf}{2}
\newcommand{\affsac}{3}
\newcommand{\affdark}{4}
\newcommand{\affnsw}{5}
\newcommand{\affarc}{6}
\newcommand{\affsifa}{7}
\newcommand{\affhawaii}{8}
\author[Stokholm et al.]{Amalie Stokholm$^{\affdifa,\affinaf,\affsac}$\thanks{amalie.stokholm@unibo.it}, Víctor Aguirre Børsen-Koch$^\affdark$, Dennis Stello$^{\affsac,\affnsw,\affarc,\affsifa}$, Marc Hon$^\affhawaii$, \newauthor Claudia Reyes$^\affnsw$
\\
$^\affdifa$Dipartimento di Fisica e Astronomia, Universit\`a degli Studi di Bologna, Via Gobetti 93/2, I-40129 Bologna, Italy \\
$^\affinaf$INAF -- Osservatorio di Astrofisica e Scienza dello Spazio di Bologna, Via Gobetti 93/3, I-40129 Bologna, Italy\\
$^\affsac$Stellar Astrophysics Centre, Department of Physics and Astronomy, Aarhus University, Ny Munkegade 120, DK-8000 Aarhus C, Denmark\\
$^\affdark$DARK, Niels Bohr Institute, University of Copenhagen, Jagtvej 128, 2200 Copenhagen, Denmark\\
$^\affnsw$School of Physics, The University of New South Wales, Sydney NSW 2052, Australia\\
$^\affarc$ARC Centre of Excellence for Astrophysics in Three Dimensions (ASTRO-3D), Australia\\
$^\affsifa$Sydney Institute for Astronomy (SIfA), School of Physics, University of Sydney, NSW 2006, Australia\\
$^\affhawaii$Institute for Astronomy, University of Hawai`i, 2680 Woodlawn Drive, Honolulu, HI 96822, USA\\
}

\date{Accepted 2023 June 19. Received 2023 January 07; in original form 2021 December 02}

\pubyear{2023}

\begin{document}
\label{firstpage}
\pagerange{\pageref{firstpage}--\pageref{lastpage}}
\maketitle

\begin{abstract}
The Milky Way has distinct structural stellar components linked to its formation and subsequent evolution, but disentangling them is nontrivial.
With the recent availability of high-quality data for a large numbers of stars in the Milky Way, it is a natural next step for research in the evolution of the Galaxy to perform automated explorations with unsupervised methods of the structures hidden in the combination of large-scale spectroscopic, astrometric, and asteroseismic data sets.
We determine precise stellar properties for \noofstars red giants, mainly spanning $2-15$~kpc in Galactocentric radii, making it the largest sample of red giants with measured asteroseismic ages available to date.
We explore the nature of different stellar structures in the Galactic disc
by using Gaussian mixture models as an unsupervised clustering method to find substructure in the combined chemical, kinematic, and age subspace. The best-fit mixture model yields four distinct physical Galactic components in the stellar disc: the thin disc, the kinematically heated thin disc, the thick disc, and the stellar halo.
We find hints of an age asymmetry between the Northern and Southern hemisphere and we measure the vertical and radial age gradient of the Galactic disc using the asteroseismic ages extended to further distances than previous studies.
\end{abstract}


%%%%%%%%%%%%%%%%%%%%%%%%%%%%%%%%%%%%%%%%%%%%%%%%%%

%%%%%%%%%%%%%%%%% BODY OF PAPER %%%%%%%%%%%%%%%%%%
% ~~~~~~~~~~~~~~~~~~~~~~~~~~~~~~~~~~~~~~~~~~~~~~~~~~~~~~~~~~~~~~~~~~~~~~~~~~~~~
% INTRODUCTION
% ~~~~~~~~~~~~~~~~~~~~~~~~~~~~~~~~~~~~~~~~~~~~~~~~~~~~~~~~~~~~~~~~~~~~~~~~~~~~~
\section{Introduction}
\label{sec:introduction}
Understanding the formation and evolution of galaxies is an important goal in astrophysics today.
The kinematics and chemical compositions of the stars in the Milky Way contain clues of the Galactic past \citep[e.g.][]{freemanblandhawthorn2002}, meaning we can use stars of different ages as time capsules to reconstruct the Galactic evolutionary history.

From star counts and studies of bulk stellar properties we know that spiral galaxies such as our own have distinct structural components. Traditionally these have been identified as the Galactic thin and thick disc, the central bulge, and the stellar halo. Each component contains stars of distinct characteristics and the components are therefore believed to be linked to different formation epochs and timescales.
These components are not isolated entities but have significant spatial overlap so disentangling them is non-trivial. 

Due to the difficulties of dating field stars from photometric or spectroscopic information, many authors have studied the stellar populations in the Milky Way using their chemistry and kinematics. 
From galaxy formation theory, we expect that galactic components consist of stars that share similarities in chemical abundances as well as orbital energies.
However, previous studies have mainly tried to define components based only on chemistry or kinematics.
Some studies subdivide the stars based on their kinematics and then study their chemical abundance distributions \citep[e.g.][]{bensby2003, nissenschuster2010, hayden2020}.
Alternatively, other studies have selected stars based on their elemental abundances and studied whether a clear division in kinematics would be predicted \citep[e.g.][]{mackereth2017}.
As shown by e.g. \citet{silvaaguirre2018} the dissection done in either chemical space or kinematics space does not reproduce the same patterns and a clear division in one space looks less structured in the other.

Recently, advances in observations made by spacecraft and ground-based facilities have significantly improved the quality and quantity of data available to study stellar populations in the Galaxy. The \gaia mission
\refereenew{\citep{gaia,gaiaedr3,gaiadr3}}
provided kinematic phase space information up to 6 dimensions for more than a billion stars in the latest data release. This has enabled significant leaps in our understanding of the dynamics of the Galaxy \citep{helmi2020} and it has challenged the classical formation picture of our Galaxy and uncovered a number of previously unknown stellar substructures
\refereenew{\citep[e.g.][]{helmi2018,belokurov2018,myeong2019,koppelman2019,Kruijssen2020, horta2021}}.

One of the major challenges in studying the Galactic past is to achieve the necessary accuracy in stellar age. Determining physical properties of stars such as mass, radius, and especially age is remarkably difficult using traditional techniques, particularly for giant stars which are the ones that allow us to probe the furthest.
In the last few decades, the study of stellar pulsations or \emph{asteroseismology} has led to a dramatic improvement in the precision with which we can measure these important properties.
Stellar pulsations are directly related to the internal structure and composition of a star and thus to the nuclear processes in the stellar interior, based on which a precise stellar age can be determined \citep{miglio2013,chaplin2014}.
The space-based photometry missions \corot \citep{baglin2006}, \kepler \citep{gilliland2010,koch2010}, and more recently \tess \citep{ricker2014} have revolutionised the study of cool stars exhibiting stochastic oscillations which are excited and damped by near-surface convection, the so-called solar-like oscillators \citep[see e.g.][for a review]{hekker2017}. The nearly-uninterrupted space-based observations have detected these pulsations in tens of thousands of stars covering different regions of the Hertzsprung-Russell Diagram.

Even if the individual oscillation frequencies cannot necessarily be resolved, two quantities that describe the overall shape of the pulsation pattern can usually be extracted: the large frequency separation \dnu and the frequency at maximum power \numax.
These values are related to the mean density \citep{ulrich1986} and surface gravity \citep{brown1991,kjeldsen1995} of the stars, respectively. They therefore provide important constraints on the internal structure of stars, useful for precise determination of stellar properties such as age.

Given the new availability of high-quality observational data and the issues arisen from manually defining components based on only kinematics or chemistry, it is a natural next step for research in the evolution of the Galaxy to perform automated explorations with unsupervised clustering methods of the structures hidden in the data sets.
In this context, unsupervised means that we are not using any prior knowledge about Galactic components when searching for structure in the data.
The approach is automated and data-driven which means that it can be used with future increases in data collection.
This also allows for a mix in dimensions, e.g.\@ using both chemical and kinematics dimensions to find substructure in the data.

This work explores the nature of the different stellar structures in the largest sample of red giant stars with measured asteroseismic ages available to date.
We derive precise and self-consistent stellar properties for the ensemble and 
use Gaussian mixture models (GMM) as an unsupervised clustering method to find substructure in the combined chemical, kinematic, and age subspace. 

% ~~~~~~~~~~~~~~~~~~~~~~~~~~~~~~~~~~~~~~~~~~~~~~~~~~~~~~~~~~~~~~~~~~~~~~~~~~~~~
% METHOD
% ~~~~~~~~~~~~~~~~~~~~~~~~~~~~~~~~~~~~~~~~~~~~~~~~~~~~~~~~~~~~~~~~~~~~~~~~~~~~~
\section{Sample and methods}
\label{sec:sample}
% Celestial sky view and Galactocentric view of the sample
\begin{figure*}
	\centering
	\includegraphics[width=\textwidth]{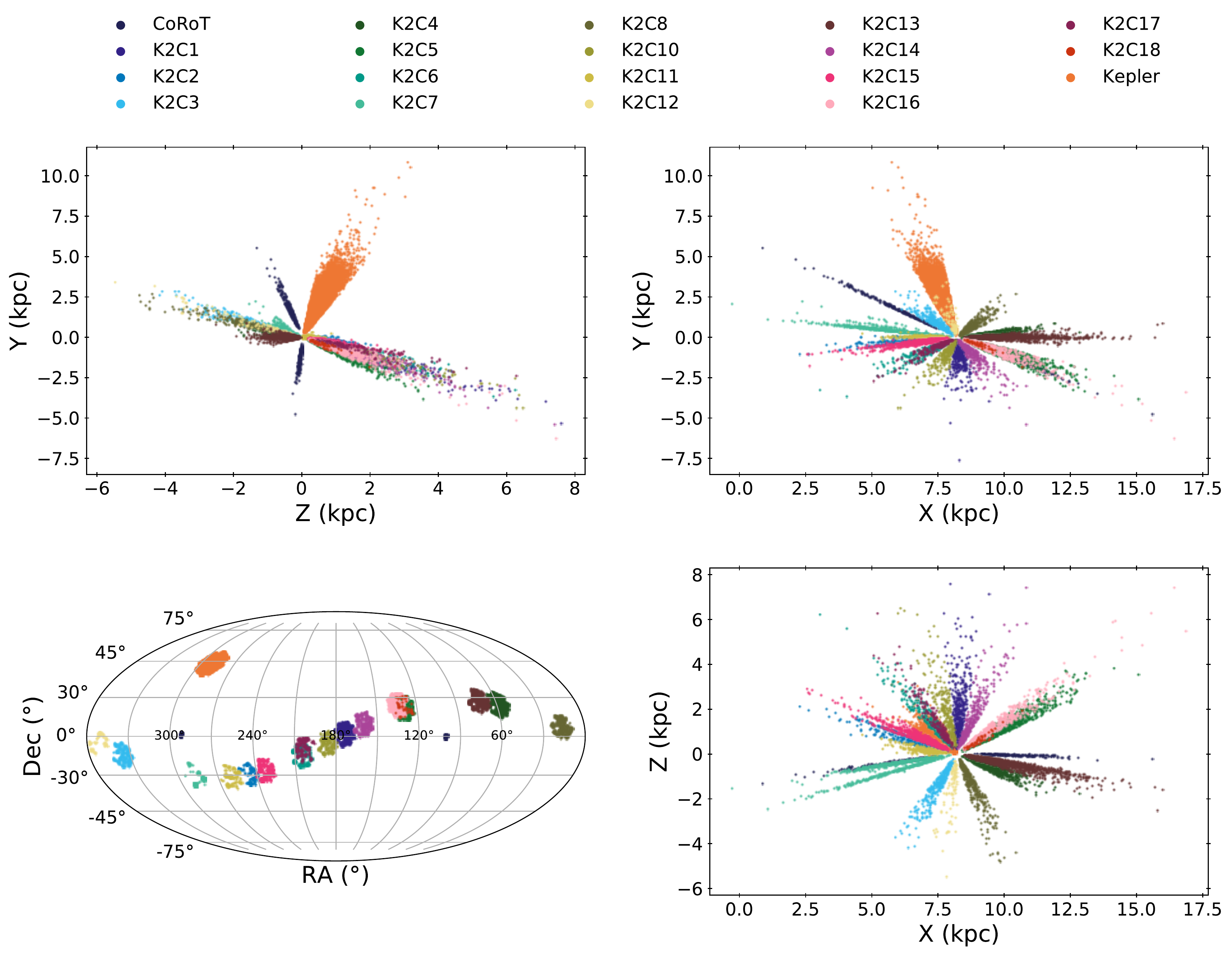}
	\caption{Spatial distribution of the stars in the final, pruned sample. The colours correspond to different surveys as indicated by the legend.
	Lower left shows a celestial overview of the sample in right ascension (RA) and declination (Dec.) in the Mollweide projection. The remaining three subfigures show the distribution of stars in Cartesian, Galactocentric coordinates in the X-Y plane (upper left), Y-Z plane (upper right), and X-Z plane(lower left) respectively.
}
	\label{fig:galacticpositions}
\end{figure*}

All stars in this study are evolved low-mass stars, either H-shell burning red giants or He-core burning red clump stars. These stars are ideal as they are relatively common and intrinsically bright, allowing us to study a representative sample of the stellar population at a great range of distances from the Sun.

\figref{fig:galacticpositions} shows the three dimensional Galactocentric positions along with the projected celestial positions of the targets in our final, pruned sample.
Our sample contains \noofstars targets with 
(i) asteroseismic quantities from time series measured by either the nominal mission of the \kepler spacecraft \citep[][\noofkeplerstars stars]{gilliland2010,koch2010},
its renaissance mission \ktwo \citep{howell2014} campaigns \ktwocn (\noofktwostars stars),
or the \corot space mission \citep[][\noofcorotstars stars]{baglin2006},
(ii) 5 dimensional astrometric parameters (positions, proper motions, and parallax) from the
early third data release of the \gaia mission \citep[\gaia EDR3,][]{gaiaedr3},
(iii) spectroscopic abundances from either
the 16th data release of the Apache Point Observatory Galactic Evolution Experiment \citep[APOGEE DR16;][\noofapogeestars stars]{apogee,ahumada2020}, 
the 3rd data release of the Galactic Archaeology with HERMES survey \citep[GALAH DR3;][\noofgalahstars stars]{buder2020}, or
the value-adding catalogue of the fifth data release from the Large Sky Area Multi-Object Fibre Spectroscopic Telescope Survey \citep[\lamost DR5;][\nooflamoststars stars]{deng2012,xiang2019},
and (iv) photometric magnitudes from the Two Micron All-Sky Survey \citep[2MASS;][]{skrutskie2006}.

\begin{figure}
    \centering
    \includegraphics[width=\columnwidth]{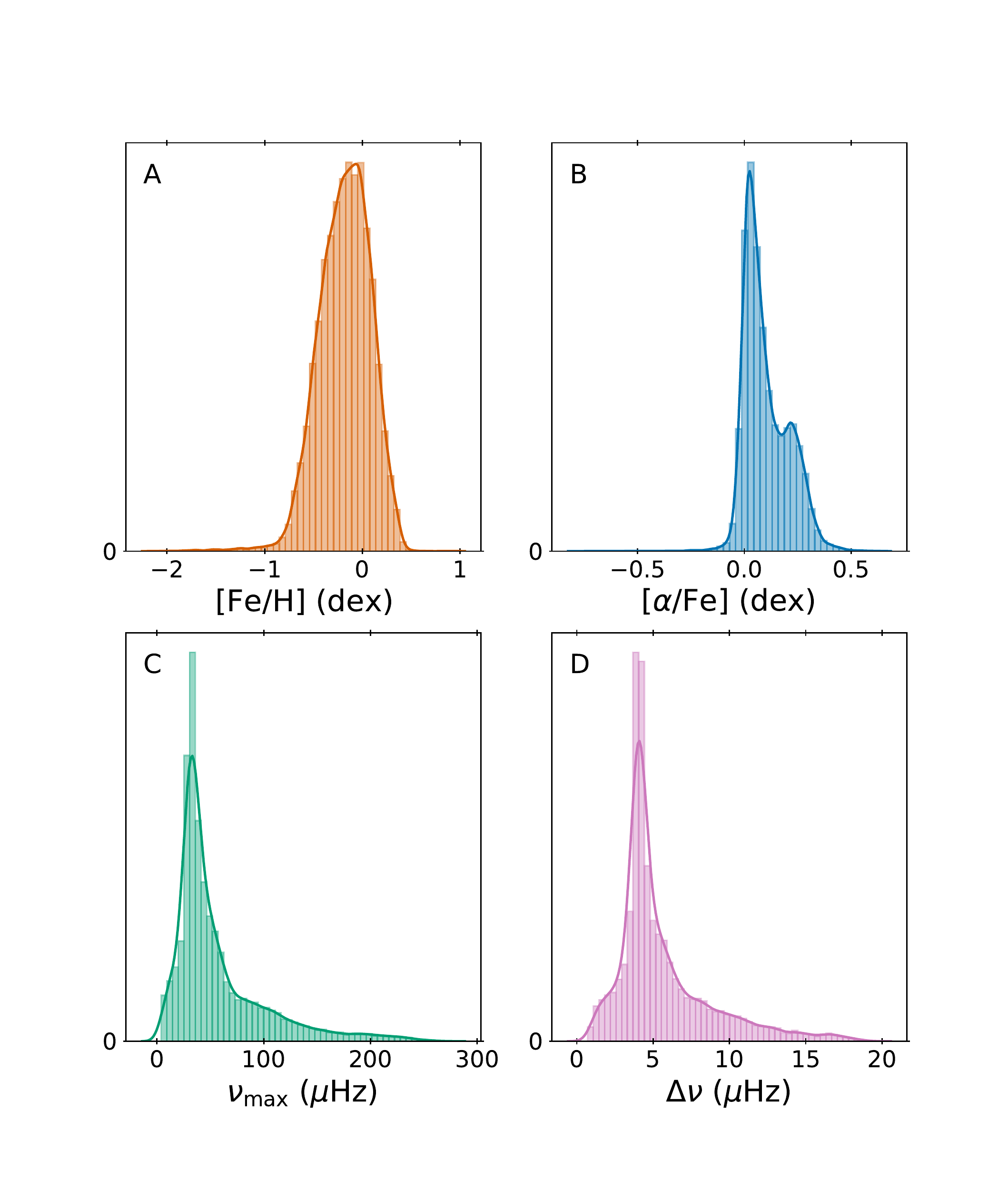}
    \caption{\refereenew{Distributions of stellar properties for the full sample of respectively (A) metallicity \feh, (B) alpha-to-iron ratio \alphafe, (C) frequency of maximum power \numax, and (D) large frequency separation \dnu.}}
    \label{fig:fourhistograms}
\end{figure}

For targets within the fixed field-of-view of the nominal \kepler mission, we use the asteroseismic quantities \numax and \dnu reported by \citet{yu2018}.
For \ktwo targets, we use the solutions for the global asteroseismic observables found from the \syd pipeline \citep{huber2009} released in the \ktwo Galactic Archaeology Program data releases \citep[K2GAP,][]{stello2017,zinn2020,zinn2021}.
For \corot targets, we use reported \dnu and \numax values from \citet{anders2017}.

No spectroscopic survey covers the full sample. Star-to-star comparisons of large-scale spectroscopic surveys show how the different abundance scales used by these different surveys can lead to significant disagreements in e.g.\@ the metallicity \citep[see e.g.][]{kunder2017, rendle2019}. We proceed with caution with this patched solution where we use spectroscopic information from different spectroscopic surveys depending on availability for a given target.
If available, we use spectroscopic data from APOGEE DR16.
If targets were not observed in APOGEE DR16, we search for spectroscopic data in either GALAH DR3 or in the value-adding catalogue of LAMOST DR5.

\begin{figure}
    \centering
    \includegraphics[width=\columnwidth]{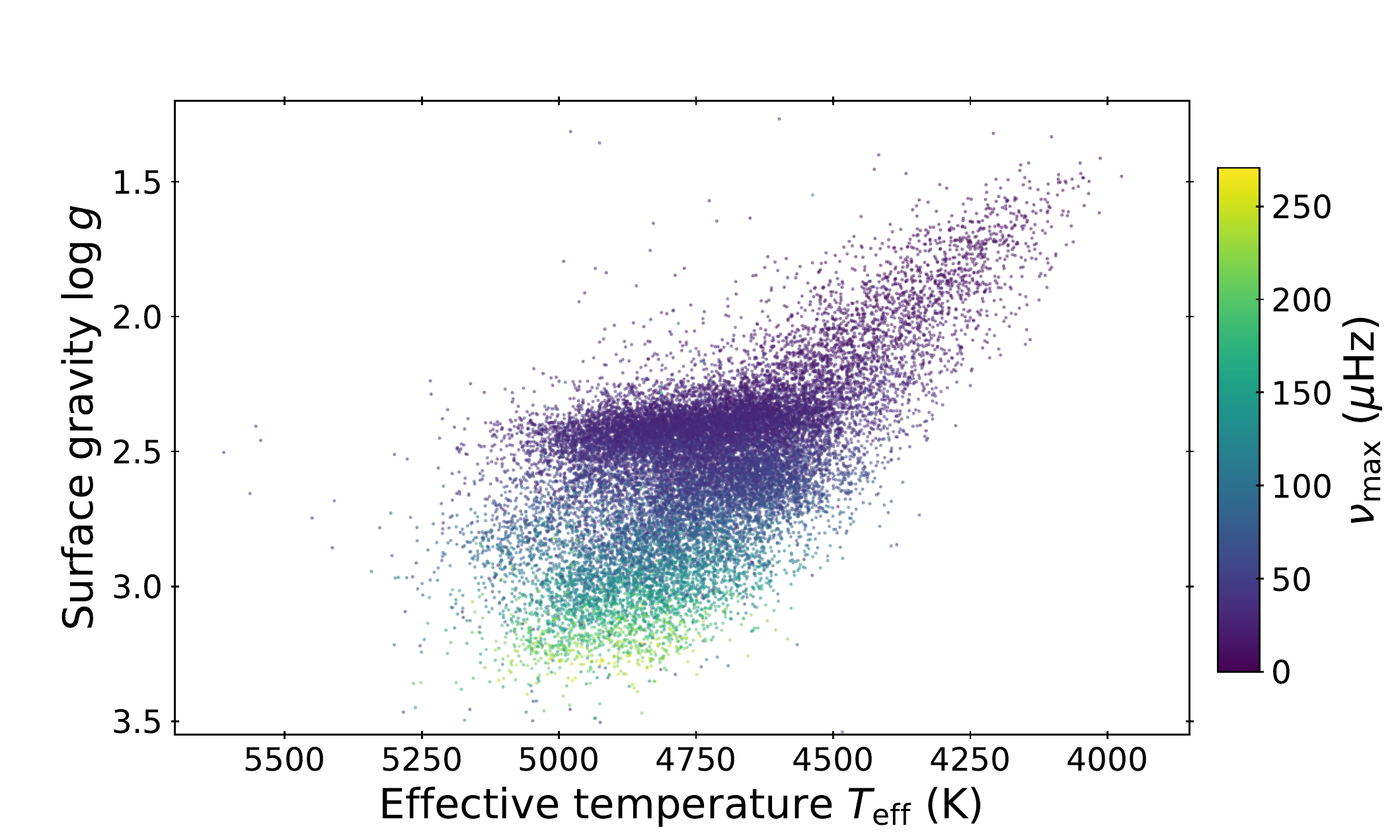}
    \caption{\refereenew{Kiel diagram of the full sample colour-coded by the frequency of maximum power \numax.}}
    \label{fig:kiel}
\end{figure}

Besides requiring that all targets have this information available, we also had the following quality criteria applied in this work:
\begin{enumerate}[i]
\item
The 2MASS quality flag \flag{Qflg}$=A$ for all bands.
\item
The K2GAP score \flag{DNU\_PROB}$>0.8$, following \citep{zinn2020}.
\item
For targets with APOGEE data: No \flag{TEFF\_BAD}, \flag{LOGG\_BAD} or \flag{VMICRO\_BAD} ASPCAP flag set%
.
\refereenew{Furthermore,}
no \flag{LOW\_SNR}, \flag{SUSPECT\_RV\_COMBINATION}, \flag{PERSIST\_HIGH}, \flag{PERSIST\_JUMP\_POS}, \flag{PERSIST\_JUMP\_NEG}, or \flag{VERY\_BRIGHT\_NEIGHBOR} star flag set.
\item
For targets with GALAH data: no \flag{flag\_sp}, \flag{flag\_fe\_h}, \flag{flag\_mg\_fe}, or \flag{flag\_si\_fe} set.
\item
For targets with LAMOST data: No \flag{teff\_flag}, \flag{logg\_flag}, \flag{vmic\_flag}, \flag{fe\_h\_flag}, \flag{mg\_fe\_flag}, \flag{si\_fe\_flag} set and \flag{qflag\_singlestar=YES}.
Furthermore, we require \flag{chi2ratio<5} and that the effective temperature should be in the range $\SI{4100}{\kelvin}<\teff<\SI{5300}{\kelvin}$ as we found systematic trends in effective temperature outside this range when comparing stars that had both APOGEE and LAMOST spectroscopic parameters available.
\end{enumerate}

Requiring that all stars have the full asteroseismic, spectroscopic, and astrometric information, the sample contains \noofstarsbeforepruning stars before quality assessment in \secref{ssec:properties}.
\refereenew{\figref{fig:fourhistograms} illustrates distributions in four of the key observables of the pruned sample, while \figref{fig:kiel} shows the stars in a Kiel diagram colour-coded by their measured \numax.}

% ~~~~~~~~~~~~~~~~~~~~~~~~~~~~~~~~~~~~~~~~~~~~~~~~~~~~~~~~~~~~~~~~~~~~~~~~~~~~~
% SYSTEMATICS
% ~~~~~~~~~~~~~~~~~~~~~~~~~~~~~~~~~~~~~~~~~~~~~~~~~~~~~~~~~~~~~~~~~~~~~~~~~~~~~
\subsection{Systematic effects}
\label{ssec:systematics}
In most catalogues, only a formal uncertainty is given.
These are typically estimated from the internal consistency of measurements and do not represent the total uncertainty on a given measurement. 
For the astrometric quantities from \gaia EDR3, we use the appropriate multiplicative unit-weight uncertainty for the parallax uncertainties determined by \citet{fabricius2021}. This factor is either 1.22 or 1.05, depending on whether the \emph{pseudocolor},
i.e.\@%
a colour estimate that minimises the residuals in the astrometric solution, was included in the determination of the solution or not.
For the uncertainty in proper motion, we follow \citet{gaiamagellanic2021} and add a systematic uncertainty of $0.01$~mas~yr$^{-1}$ for each component of proper motion. This value is of the same order as the spatially dependent systematic errors found in \citet{lindegren2021b}.
A different systematic effect to take into account is the parallax zero-point offset. \gaia EDR3 has a global parallax zero point offset of $-17$ $\mu$as when comparing to the distant quasars \citep{lindegren2021b} which we correct for.

In the LAMOST value-adding catalogue they derive the spectroscopic quantities from LAMOST spectra using a data-driven method, \emph{the Payne} \citep{ting2019}, trained on either APOGEE or GALAH data. Targets with data from this catalogue thus inherit systematics effects from the training data. We take this into account in the same fashion as in \citet{xiang2019} and add the systematic effects found from external validation \citep{ting2019} in quadrature to the formal uncertainties.

For all asteroseismic measurements, we estimate the systematic uncertainties from the scale factors computed in \citet{zinn2021},
designed to put the average \numax and \dnu values from the different light curve pre-processing on the same scale. We compute the standard deviation of these scale factors for both \dnu and \numax for stars in the red clump phase and stars on the red giant branch separately and use these scatters as an additive fractional systematic uncertainty. These systematic uncertainties are less than 1 percent for both \numax and \dnu. For stars with no evolutionary information, we use the greater of the two estimates for each parameter.

% ~~~~~~~~~~~~~~~~~~~~~~~~~~~~~~~~~~~~~~~~~~~~~~~~~~~~~~~~~~~~~~~~~~~~~~~~~~~~~
% BASTA
% ~~~~~~~~~~~~~~~~~~~~~~~~~~~~~~~~~~~~~~~~~~~~~~~~~~~~~~~~~~~~~~~~~~~~~~~~~~~~~
\subsection{Determination of stellar properties}
\label{ssec:properties}
% BASTA
We determine the stellar properties such as mass, radius, and age for all stars in our sample using the BAyesian STellar Algorithm \citep[\basta;][]{silvaaguirre2015,silvaaguirre2017, aguirreborsenkoch2021}.
\basta uses Bayesian inference to determine the probability density function of a stellar parameter given the observed properties and other prior information and given a set of the stellar models. As the latter, we used the updated BaSTI (a Bag of Stellar Tracks and Isochrones) stellar models and isochrones library \citep{Hidalgo:2018dy,pietrinferni2021}.
We use their so-called \emph{best physics scenario} for these isochrones, which includes main sequence convective core overshooting, atomic diffusion of He and metals, and mass loss using the \citep{reimers1975} prescription and an efficiency of $\eta=0.3$.

For solar values, we adopt $\nu_{\text{max, }\odot}=3090$~$\mu$Hz, $\Delta\nu_{\odot}=135.1$~$\mu$Hz \citep{huber2011}, and $T_{\text{eff, }\odot}=5777$~K.
Theoretical predictions of \dnu and \numax are computed using the asteroseismic scaling relation along an evolutionary track or isochrone, however, the accuracy of the asteroseismic scaling relations across different metallicities, effective temperatures, and evolutionary status is currently an active discussion within the field of asteroseismology \citep[for further discussion see e.g.][]{white2011,belkacem2011,sharma2016,viani2017}. We correct the large frequency separation \dnu scaling relation following \cite{serenelli2017} as it reproduce the results for a number of classical age determinations such as the open clusters M67 \citep{stello2016} and NGC~6819 \citep{casagrande2016}.
\basta allows the possibility to add prior knowledge to the Bayesian fit, and we used the Salpeter Initial Mass Function \citep{salpeter1955} as a prior to quantify our expectation of most stars being low-mass stars.
The BaSTI isochrones cover ages up to $14.5$~Gyr so we include this as an upper limit on stellar age.

\begin{figure}
    \centering
    \includegraphics[width=\columnwidth]{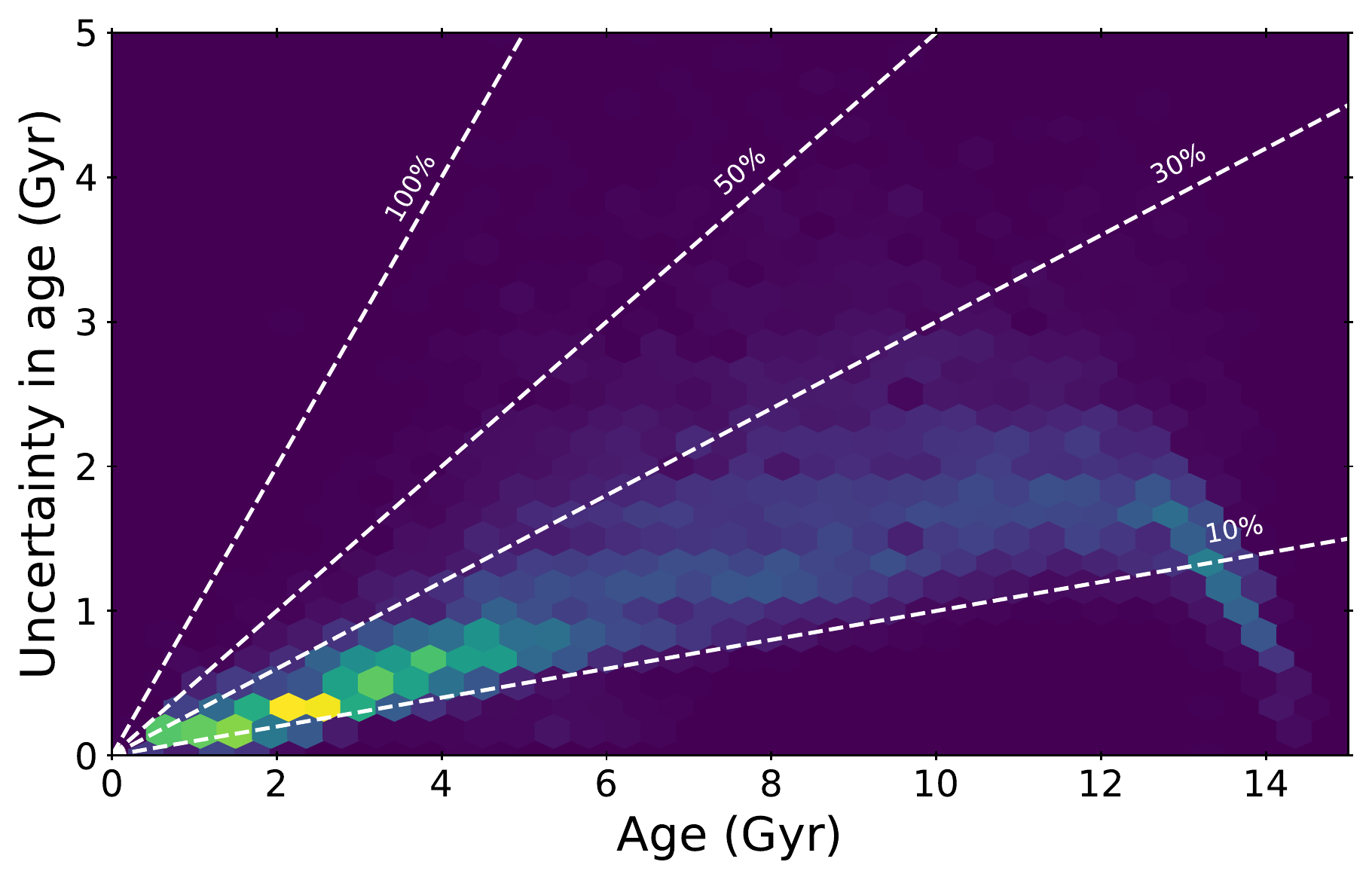}
    \caption{\refereenew{Age vs.\@ uncertainty in age in Gyr. As we are working with the probability distributions and thus allowing asymmetric upper and lower uncertainties, the uncertainty used here is a simple mean of the upper and lower uncertainty estimates as defined in \secref{ssec:properties}. Note how this value drops near oldest stars due to edge effects as the posterior distribution is limited to be within the age range set by the grid of stellar models.}}
    \label{fig:ageuncertainty}
\end{figure}

We fit the spectroscopic properties effective temperature, \teff, and metallicity \meh, the asteroseismic values \dnu and \numax, the photometric apparent magnitudes in the 2MASS bands $JHK_s$, and the corrected astrometric parallax \parallax.
For \knownphase stars in the sample, the evolutionary state of the star had been determined from asteroseismology \citep{bedding2011,stello2013,mosser2014,vrard2016,hon2017,elsworth2017,zinn2020} and this phase information was also added as a constraint. 
The metallicity estimates \meh are computed following the formalism from \citet{salaris1993} as $\meh = \feh + 0.694 \times 10^{\alphafe} + 3.06$, while the used \alphafe abundances are computed as [(Mg + Si)/Fe]. Mg and Si are the two $\alpha$-elements that contribute most significantly to the opacities in this temperature range and this estimate corresponds most closely to the \alphafe in stellar evolution models \citep{vandenberg2012,salaris2018}.
We remove \noofremovedbasta stars where the difference between the input and output value is more than three standard deviation in either \teff, \dnu, or \numax.
We also fitted the sample using only the spectroscopic and asteroseismic quantities and we remove stars from the sample if the radius from the two different solutions differ by more than three standard deviations.
The derived stellar parameters for this final sample of \noofstars stars can be found in Table~\ref{tab:cat}.
If we let the mean of the 16th and 84th quantiles be a measure of symmetric uncertainty, the median fractional uncertainty in age is \SI{15}{\percentage} for the entire sample. \january{This is similar to median fractional uncertainties in age of other studies using asteroseismic quantities \citep[e.g.][]{anders2017,silvaaguirre2018,miglio2021} and it shows how especially the inclusion of the evolutionary phase for a majority of the stars in our sample in combination with the other asteroseismic observables narrows down the likely parameter space and thus the statistical uncertainties.}
\refereenew{The behaviour of this estimated symmetric uncertainty as a function of stellar age can be seen in \figref{fig:ageuncertainty}.} 
\january{\citet{miglio2021} notes how the stellar age scales with \dnu to the power of $\sim14$, which translates to a higher resolution at younger ages than at older ages, a trend that is also visible in our sample. \figref{fig:ageuncertainty} also shows a downturn in mean uncertainty at $12$~Gyr, which is attributed to edge effects within the grid of stellar models. The posterior distribution of a given star is limited to be within the age range set by the grid so for a star nearing the upper limit of the grid in age, the posterior becomes quite asymmetric as the upper uncertainty decreases due to these limitations set by the model grid.}
\refereenew{Some of these stars have ages available from studies that did not include the asteroseismic information in their derivation.
One set of available ages are from the astrophysical parameters inference system (Apsis) published in the third data release of \gaia \citep{fouesneau2022} which combines astrometric data with photometric and spectroscopic measurements from \gaia. Even if we only consider the stars with good measurements (\texttt{flag\_flame=00}), their median fractional uncertainty in stellar age is above \SI{50}{\percentage} and their ages were found to be significantly smaller than the ages determined in this study. As many of these stars have similar photometric properties, this kind of age determination suffers significant degeneracies in the stellar properties inference. These degeneracies are partly lifted when including the asteroseismic measurements.}

\begin{figure}
	\centering
	\includegraphics[width=0.9\columnwidth]{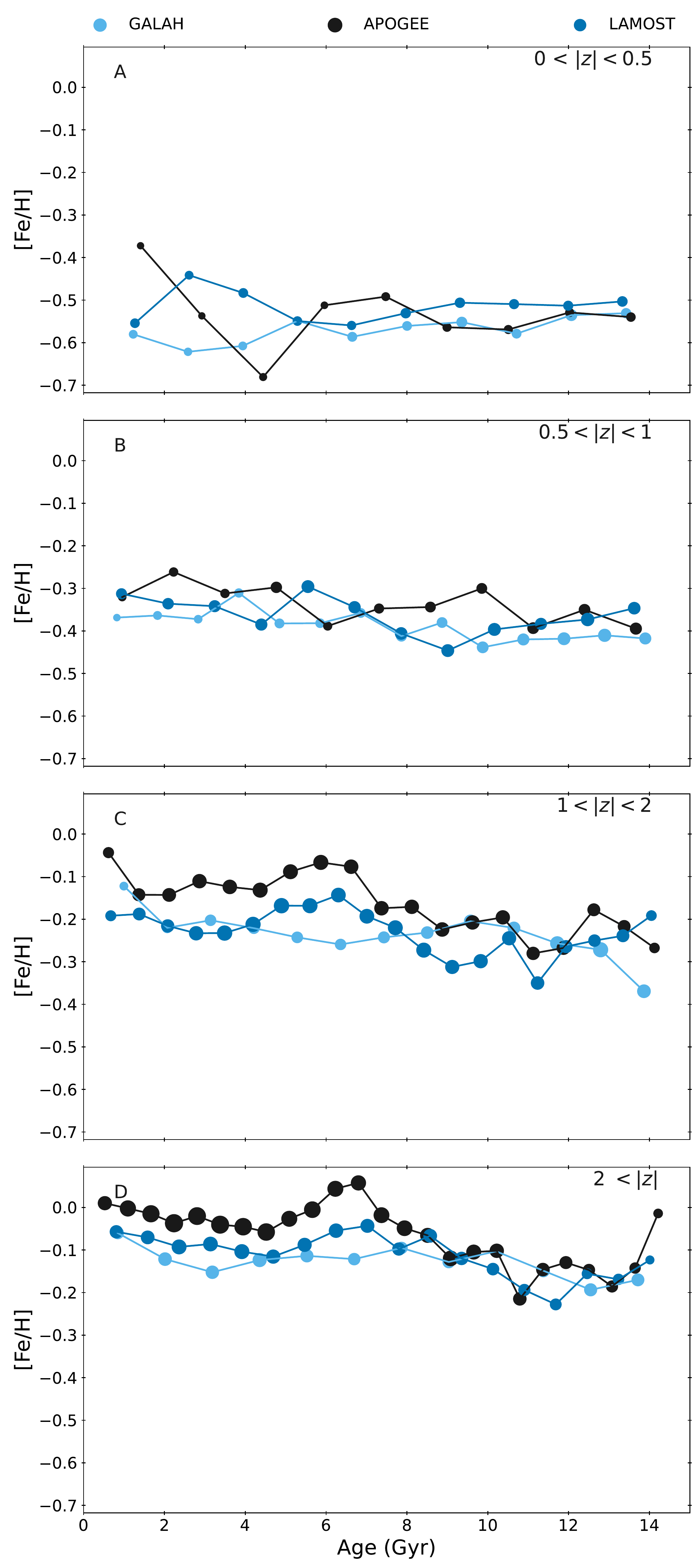}
	\caption{\refereenew{The median age vs.\@ metallicity plane split into different slices in distance away from the Galactic midplane $|z|$, colour-coded by the spectroscopic survey that determined the \feh. The bin sizes were determined using the Freedman Diaconis rule and the sizes of the points scale with the number of targets in each bin.}}
	\label{fig:agemetallicityzgcsplit}
\end{figure}
\refereenew{In \figref{fig:agemetallicityzgcsplit}, the age vs.\@ metallicity plane is shown for targets at different distances away from the Galactic midplane $|z|$ and coloured by which spectroscopic survey determined the metallicity of the targets. Even though these targets are found along different pointings in the two other Galactic coordinates $\galr$ and $\phi$ -- which should cause variation in this plane -- they overall roughly follow the same trends with age, showing that the ages are more or less consistent in the age-metallicity plane between surveys for stars at the same height.}

% ~~~~~~~~~~~~~~~~~~~~~~~~~~~~~~~~~~~~~~~~~~~~~~~~~~~~~~~~~~~~~~~~~~~~~~~~~~~~~
% ORBITS
% ~~~~~~~~~~~~~~~~~~~~~~~~~~~~~~~~~~~~~~~~~~~~~~~~~~~~~~~~~~~~~~~~~~~~~~~~~~~~~
\subsection{Orbits}
% Kinematics 
Stars born in the Milky Way within the last \SI{8}{Gyr} were presumably born on disc-like orbits. However, the stellar discs are vulnerable to gravitational perturbations, which causes stars to migrate and diffuse from their birth orbits.
% Action-angles
Action-angles ($J$, $\theta$) are the canonical coordinates to describe stellar orbits. For an axisymmetric gravitational potential, all three actions (\Jr, \Jphi, \Jz) are integrals for motions, where \Jr describes the oscillations of a orbit in the radial direction, \Jz the oscillations in the vertical direction, and \Jphi is the azimuthal action, which is equal to the angular momentum in the vertical direction \Lz. These all share the same units \si{\kilo \parsec \times \kilo\metre\per\second} and in contrast to the coordinates in position or velocity space, actions are constant over time.
The three corresponding angles (\thetar, \thetaphi, \thetaz) quantifies the orbital phase and evolve linearly with time, modulo $2\pi$.
In a non-axisymmetric potential, the actions are not well defined and not exactly integrals of motions. The Milky Way has non-axisymmetric features such as its spiral arms and bar, however, axisymmetric approximations can still be applied and make it possible to compute these canonical coordinates.
For a detailed description of action angle variables, we refer the reader to \citet{binneytremaine2008}.

The orbital properties of the stars were computed from the 5D astrometric information from \gaia EDR3 and along with line-of-sight velocities from preferably the same spectroscopic survey that provided the chemical abundances of a target, otherwise from \gaia EDR3.
For \ruwepercent~per cent of our final sample, the re-normalised unit weighted error was less than 1.4 \citep{ruwe}, and for \ruwepercentsingle~per cent was less than 1.2.
We use the \textsc{Python} package \texttt{galpy} \citep{galpy} for the coordinate and velocity transformation. We use the implementation of the action-angle estimation algorithm \emph{St\"ackel fudge} \citep{binney2012} in \texttt{galpy} with a focal length focus of $0.45$ to calculate orbit information such as actions, eccentricity, and maximum orbit Galactocentric height.
For the calculation of orbit information,
we use the \texttt{MWPotential2014} axisymmetric gravitational potential with a circular velocity of \SI{240}{\kilo\metre\per\second} at the solar radius of \SI{8.2}{\kilo\parsec} \citep{reid2014,gravity2019}.
For the Galactic location and velocity of the Sun, we assume 
$(X,Y,Z)=(8.2,0,0.0208)$~\si{\kilo pc} and
$(U,V,W)=(11.1,12.24,7.25)$~\si{\kilo\metre\per\second} \citep{schonrich2010,bennett2019}.
By using the uncertainties in and correlation between the astrometric measurements to construct a multivariate normal distribution we run 10.000 iterations and build distributions of the orbital properties. We then use the 16th, 50th, and 84th quantiles of these distributions as a numerical estimate of the orbital values and corresponding uncertainties.

\section{Geometric dissection of the disc}
\label{sec:analysis}
If we dissect the sample into different bins in Galactocentric radius $R$ and height above the Galactic disc plane $z$, we gain insight into the geometric distribution of different stellar properties.

% ~~~~~~~~~~~~~~~~~~~~~~~~~~~~~~~~~~~~~~~~~~~~~~~~~~~~~~~~~~~~~~~~~~~~~~~~~~~~~
% RADIAL AGE GRADIENT
% ~~~~~~~~~~~~~~~~~~~~~~~~~~~~~~~~~~~~~~~~~~~~~~~~~~~~~~~~~~~~~~~~~~~~~~~~~~~~~
\begin{figure}
	\centering
	\includegraphics[width=\columnwidth]{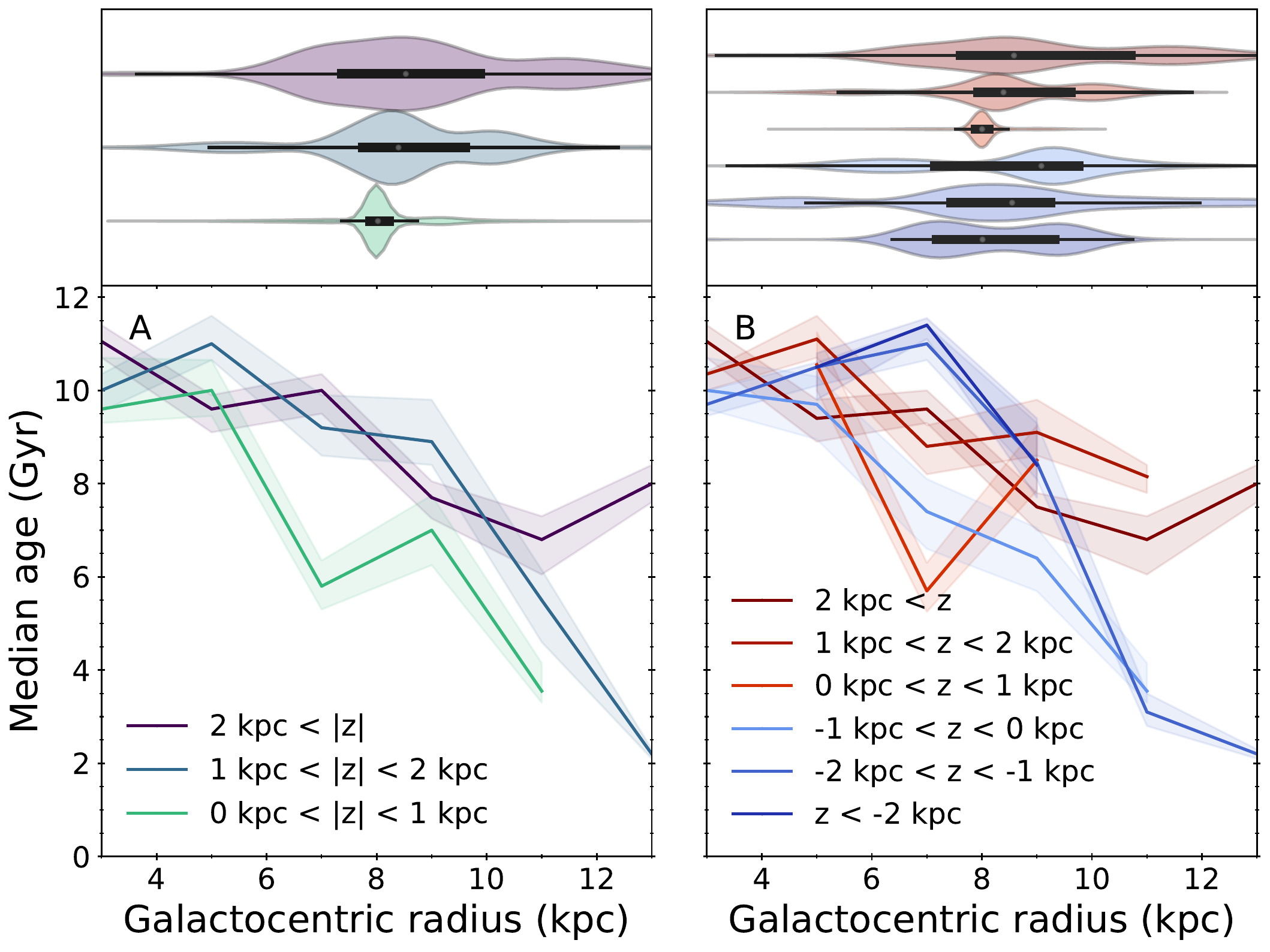}
	\caption{Radial profiles of age for the entire sample at heights $z$ from the Galactic mid-plane. The solid lines represents the median values of the bins and the shaded regions mark the area between the 16th and 84th percentiles of 1000 bootstrap samples of each bin. The bin width is $2$~kpc in \galr and there is at minimum of 10 stars in each bin, which causes a difference in extent between different slices in $z$. \refereenew{The upper panels show the normalised distributions of the stars in each selection using the same colour-code.}}
	\label{fig:radialgradient}
\end{figure}

\begin{figure}
	\centering
	\includegraphics[width=\columnwidth]{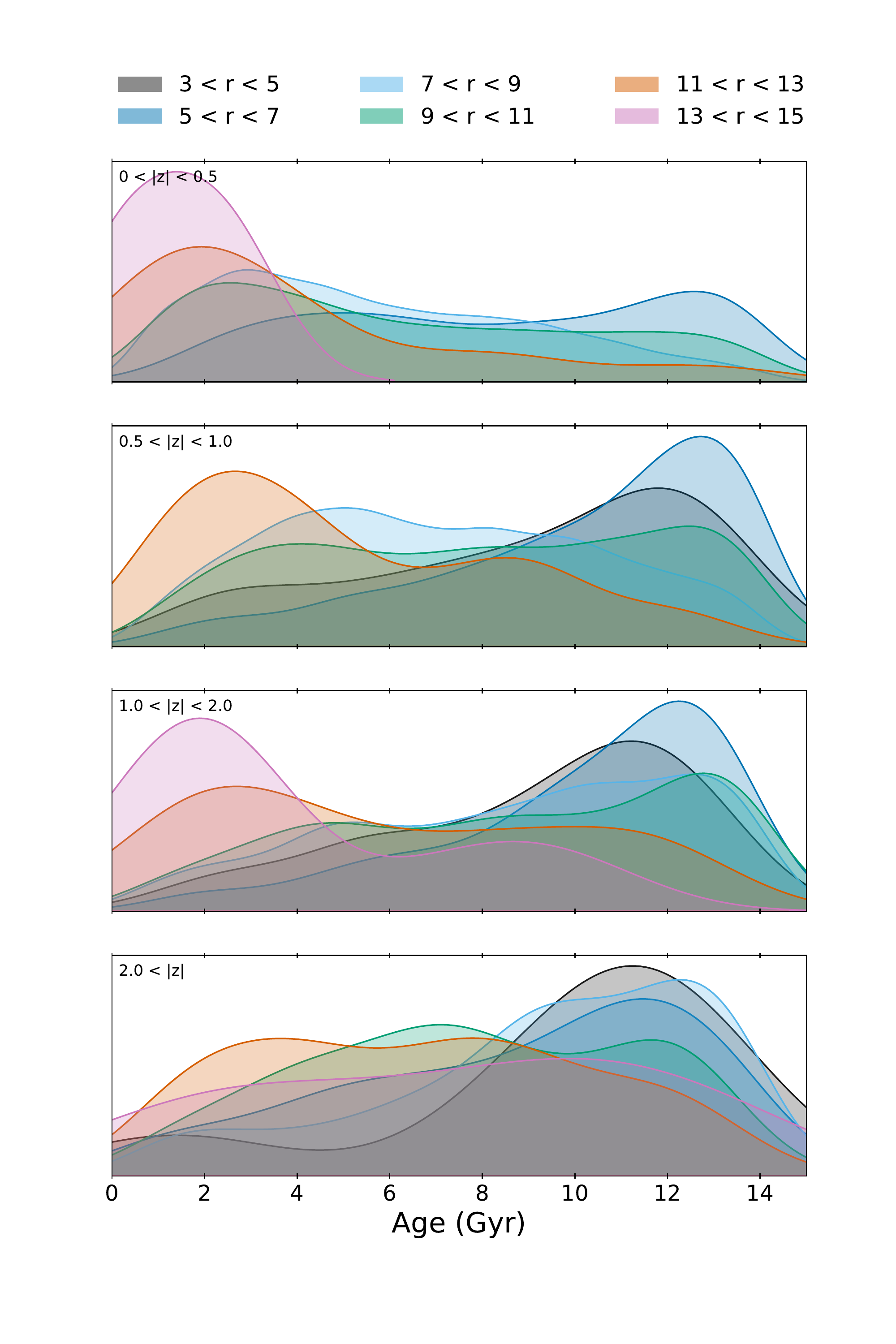}
	\caption{\refereenew{Kernel density estimates showing the distributions in age for different selection in Galactocentric radii \galr and height away from the Galactic midplane $|z|$.}}
	\label{fig:agehistograms}
\end{figure}

The \figref{fig:radialgradient}A shows the stellar age of the sample at different Galactocentric radii, split into slices in distances to the Galactic mid-plane $|z|$. We generate 1000 bootstrap draws in order to use the 16th and 84th percentiles as an estimate of the uncertainty on the median in each bin \refereenew{and the normalised distributions of each selection is shown for reference above the plot.}
We find a declining trend in age as we move outward in the Galaxy with a slope of about $-0.80$~Gyr / kpc for stars $|z| < 1$~kpc, and we find the same overall, decreasing trend across the different height slices.
These gradients are consistent with the classical \emph{inside-out} radial growth of a galactic disc, meaning that star formation sets in earlier in the more dense central regions of the galaxy than in the outskirts of it.
The idea of inside-out galactic evolution is supported by observations of external galaxies \citep[e.g.][]{ferguson2004,vandokkum2013,pezzulli2015,rodriguezpuebla2017} and cosmological zoom-in simulations \citep[e.g.][]{grand2017} as well as models of gas settling and star formation \citep{larson1976, somerville2008}.
Simulations of mono-age populations \citep{bird2012,martig2014,minchev2015} find that older stellar populations have smaller scale-lengths and larger scale-heights than younger population, meaning that younger stellar populations form in disc-like configurations that gets increasingly thinner and more extended as time goes on. As a consequence, we expect there to be a radial age gradient in the Galactic disc, with older stars at smaller Galactocentric radii and younger stars at larger radii as we see in \figref{fig:radialgradient}.
\refereenew{Another way of visualising these trends can be found in \figref{fig:agehistograms}. Here ts might be more visible how the outer regions of the Galactic disc reach younger ages, while the inner parts are predominantly older.}

The radial age gradient in \figref{fig:radialgradient}A gets shifted slightly towards higher ages as we move away from the Galactic plane.
Above $z\sim2$~kpc we see that the median age starts to increase in the outer regions, which could indicate that the flaring of the young disc stars in these regions does not reach these heights.

% Asymmetry
To further examine this effect, in \figref{fig:radialgradient}B the sample is split into bins in Galactocentric height $z$ and the median age was again computed as a function of Galactocentric radii in $2$~kpc-wide bins. However, in contrast to the \figref{fig:radialgradient}A, we now differentiate between stars above and below the Galactic mid-plane. 
The stars above and below the plane share characteristics in the innermost part of the Galaxy, but the stars above $1$~kpc from the plane reach greater median ages in the outermost part of the disc, while the median ages of stars below the plane continues to drop.
If we separate the sample by evolutionary phase and look at red clump stars and red giant branch stars separately, we still find the same trends.
Outside a Galactocentric radii of $\galr=10$~kpc, there are 1317 stars in our sample, distributed as $548$ stars above the Galactic mid-plane with $z > 0$ (red colours in \figref{fig:radialgradient}) and $769$ below the plane $z < 0$ (blue colours in \figref{fig:radialgradient}), so the numbers are comparable. As we see in \figref{fig:radialgradient}, we do probe closer to the disk below the plane than above, but stars at $1 < |z| < 2$~kpc above and below the disc still behave differently with stars above the plane being older than below.
This asymmetry in age between the two hemispheres could be explained if we probe different substructures above and below the disc, but we have not found any of such that correlates within this subset of stars.
When we inferred the stellar ages, there was a slight dependence on the positions of the stars as \basta uses a dust map, currently {\tt Bayestar19} \citep[][]{green2019}, when including the measurements of apparent magnitudes and parallax in the Bayesian inference. However, even when we use the set of the stellar ages computed without the constraints from astrometry and photometry, this asymmetry in age between the two hemisphere is still present.
\citet{hon2021} compute the masses for red giants observed with TESS, and as a tight relation between age and mass exists for red giant stars, we compare our study to theirs.
Their Figure 12 might also hint at a small age asymmetry between the two Galactic hemispheres with more massive and thus younger stars being more predominant below the Galactic plane.
This is something that motivates further study of the ages of stars in the outer region of the Galaxy.

Our measurement of the radial age gradient and how it changes as we move away from the Galactic mid-plane on \figref{fig:radialgradient}A are similar to the results of \citet{martig2016}.
At the solar radius, the median age of our sample of about $6.5$~Gyr near the disc plane and it increases to about $9$~Gyr at heights of $1$~kpc, in agreement with the vertical age gradient of $4\pm2$~Gyr~kpc$^{-1}$ measured by \citet{casagrande2016} from similar stars at the solar radius.

% ~~~~~~~~~~~~~~~~~~~~~~~~~~~~~~~~~~~~~~~~~~~~~~~~~~~~~~~~~~~~~~~~~~~~~~~~~~~~~
% VERTICAL AGE GRADIENT
% ~~~~~~~~~~~~~~~~~~~~~~~~~~~~~~~~~~~~~~~~~~~~~~~~~~~~~~~~~~~~~~~~~~~~~~~~~~~~~
\begin{figure}
	\centering
	\includegraphics[width=\columnwidth]{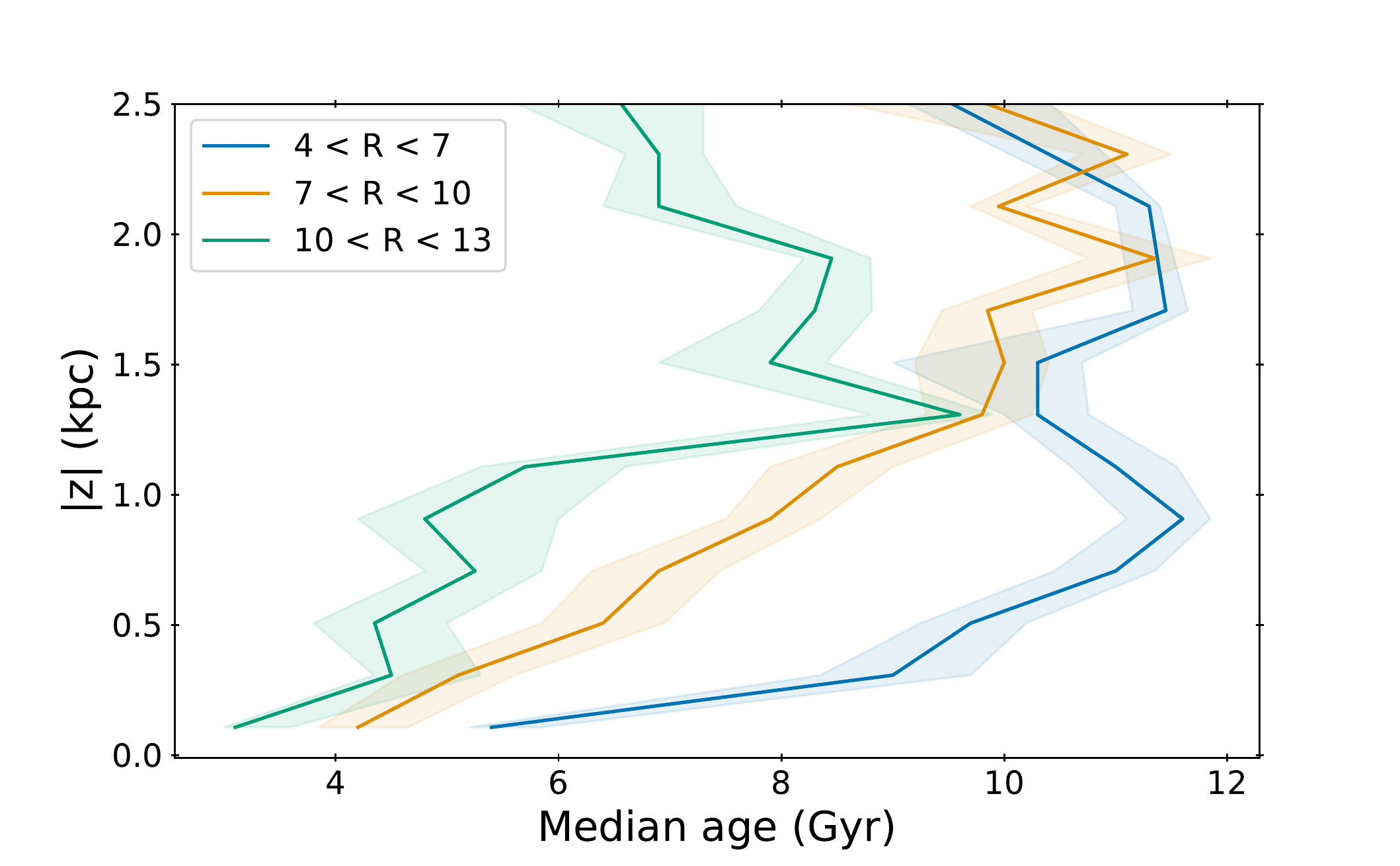}
	\caption{Vertical profiles of age for the sample between at current Galactocentric radii of $\galr=4$-$13$~kpc in bins of $3$~kpc.}
	\label{fig:verticalgradient}
\end{figure}
We compute an estimate of the vertical age gradients using stars with $\galr=4$-$13$~kpc as seen in \figref{fig:verticalgradient}, separating them into the inner Galactic disc ($4<R<7$ kpc), the Solar neighbourhood ($7<R<10$ kpc), and the outer Galactic disc ($10<R<13$ kpc).
The vertical age gradient is not expected to be fully linear as a function of the height above the disc as there is an upper limit on the ages of stars and as we are studying multiple stellar populations.
In the outermost regions and in the Solar neighbourhood, there seems to be an increasing trend with age for increasing $|z|$ up to ages of about $10$~Gyr or $|z|\sim1.5$. If we only look at the stars closest to the Galactic mid-plane $|z| < 1$~kpc, the trend seems approximately linear with a slope of $0.21$~kpc/Gyr in the Solar neighbourhood, while the slope of the outer region is slightly steeper as a function of height $|z|$ with a slope of $0.32$~kpc/Gyr. This trend however is almost non-existent in the inner regions, where the median age of about $10$~Gyr is reached below a height of $|z|\sim0.5$. This is in agreement with inside-out formation of the disc, where older populations should dominate the inner part and younger stars the outer part.
\citet{mackereth2017} find
that mono-age populations flare with age, meaning that the thickness of the disc population increase with radius, with the flaring radius decreasing with age so that the youngest populations flare the most.
This increasing flaring for younger stars can explain why the stars in the outer region stay young even at greater $z$.
For the outer region, there seem to be a sharper transition in median age around a height of $|z|\sim1.3$~kpc away from the plane. This is caused by whether it is stars above or below the plane that dominate the bin in $z$. If we compare to \figref{fig:radialgradient}, we see that in this outer region the stars at $|z|<1$ ($|z|>1$) are mostly places below (above) the Galactic mid-plane. This change in whether most stars originate from above or below the plane thus causes the rather significant shift seen for the green curve around $|z|\sim1.3$~kpc in \figref{fig:verticalgradient}.
Overall, these measurements of the vertical age gradient in the Galaxy agrees well with the vertical age gradient measured by \citet{casagrande2016}, however, our sample extends to larger distances and highlights its radial dependence.

 \subsection{Age-metallicity relations}
\refereenew{The presence of a correlation between stellar age and metallicity is a basic prediction of galactic evolutionary models; as stars evolve they enrich the galaxy with heavier elements meaning that successive generations of stars are more metal-rich.
We find a weak correlation in the age-metallicity plane with a significant scatter in metallicity at all ages, with a relatively large scatter in metallicity up to stellar ages of $8$--$9$~Gyr after which the scatter in metallicity increases significantly.
This is in agreement with previous studies of the age-metallicity relation in the Milky Way disc \citep[e.g.][]{edvardsson1993,feltzing2001,casagrande2011,silvaaguirre2018, feuillet2018}, and suggest that the chemical evolution of the Milky Way was more complicated than a simple closed box model.}

\begin{figure}
    \centering
    \includegraphics[width=\columnwidth]{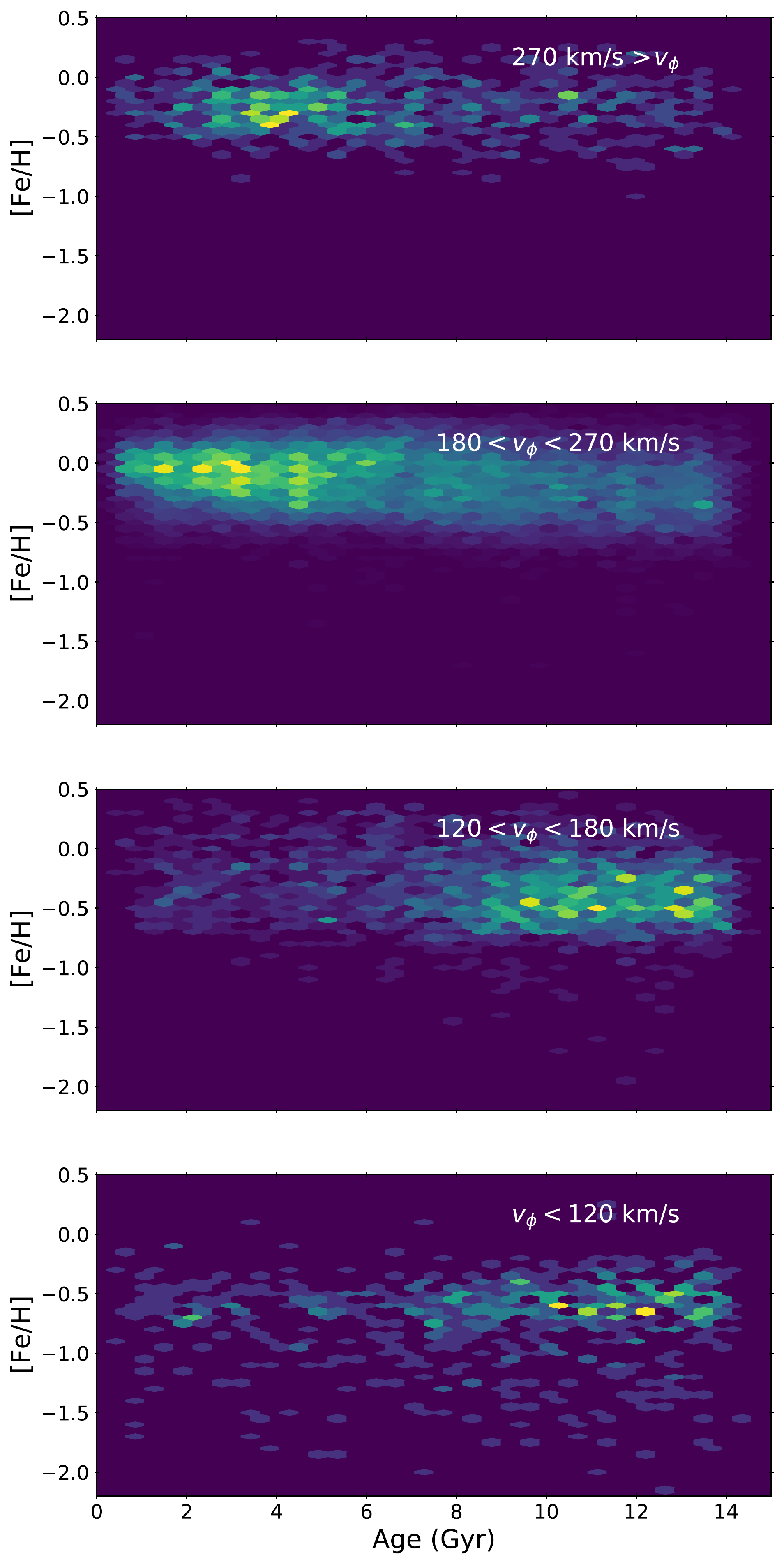}
    \caption{\refereenew{The age-metallicity plane plotted for four different subdivision of the sample according to the tangential velocities of the stars.} \january{Panel A shows the values of $762$ stars, B $17,552$, C $2183$, and D $579$.}}
    \label{fig:orbitvelocityagemetallicity}
\end{figure}

\refereenew{We study the age-metallicity in more detail by subdividing the stars into bins based on their kinematics. We choose to make the same subdivision as the study of main-sequence and subgiants stars of \citet{sahlholdt2022} in tangential velocity based on the trends found in the metallicity-tangential velocity plane. The results of this subdivision can be seen in \figref{fig:orbitvelocityagemetallicity}.}

\refereenew{For the lowest velocities in panel D, we see a low-metallicity peak between $8$--$14$~Gyr. Stars in this subdivision are mostly halo stars, so this is in agreement with other studies of the age distribution of halo stars \citep[e.g.][]{gallert2019}. However, the relation does extend down to low ages and remains rather flat.
We also see a lot of structure in age. Some of this could be attributed to a number of seemingly young \alphafe-rich stars which could be stellar merger remnants and thus look younger than they are \citep[][]{yong2016,jofre2016,silvaaguirre2018}. It could also suggest that not all halo-like structures in the Galaxy are equally old, which is the focus of a future study.}

\refereenew{In panel C we see the stars with tangential velocities between $120$--$180$\si{\kilo\metre\per\second}. While we do see a resemblance of the behaviour of the lowest velocity stars with the increased scatter at high ages, we mostly see a flat relation. Compared to \citet{sahlholdt2022}, we do not see a tight, decreasing age-metallicity relation with stellar age for the oldest stars in this group.}

\refereenew{In panel A and B, we see the majority of our sample as most of the stars in our sample have velocities similar to that of the Sun. We see a shift towards younger ages compared to panel C and D but still a flat relationship between age and metallicity. For the high velocity stars in panel A there is a peak in the distributions around an age of $4$~Gyr and metallicity of $-0.35$~dex. For the stars in panel B, a similar peak exists but it is broader in both dimensions and moved towards solar metallicity as expected.}

% ~~~~~~~~~~~~~~~~~~~~~~~~~~~~~~~~~~~~~~~~~~~~~~~~~~~~~~~~~~~~~~~~~~~~~~~~~~~~~
% DISC
% ~~~~~~~~~~~~~~~~~~~~~~~~~~~~~~~~~~~~~~~~~~~~~~~~~~~~~~~~~~~~~~~~~~~~~~~~~~~~~
 \section{Disc components}
 
% ~~~~~~~~~~~~~~~~~~~~~~~~~~~~~~~~~~~~~~~~~~~~~~~~~~~~~~~~~~~~~~~~~~~~~~~~~~~~~
% GMM
% ~~~~~~~~~~~~~~~~~~~~~~~~~~~~~~~~~~~~~~~~~~~~~~~~~~~~~~~~~~~~~~~~~~~~~~~~~~~~~
 \subsection{The Gaussian Mixture Model}
To find and quantify components in the data we use Gaussian mixture models (GMM) as a clustering algorithm. A clustering algorithm takes as input $n$ points in $D$ dimensional space and computes a clustering that assigns each point to one of \nc clusters. For GMM, the number of clusters \nc is a hyperparameter specified in advance but below we describe how we determine a good value for \nc.

A Gaussian mixture model (GMM) is a probabilistic model that consists of \nc multivariate Gaussian densities with mean \gmean, covariance \gcov, and a prior probability \gweight. Formally, a GMM is the probability density function
\begin{equation}
    p(\data,\error \mid \hyper) = \sum_{i=1}^{\nc} \gweight g(\data \mid \gmean, \gcov)
\end{equation}
over $D$-dimensional data \data,
that is, a weighted sum of Gaussian densities $g(\data \mid \gmean, \gcov)$.
Since $p(\data\mid\hyper)$ should be a probability density function, it is required that the mixture weights \gweight sum to unity. As the complete Gaussian mixture model is parameterized using the mean vectors, covariance matrices, and mixture weights from all component densities, we can collectively denote them by $\hyper = \{\gweight, \gmean, \gcov \mid i=1,\dots,\nc\}$.
The GMM assigns a point $x$ to the cluster $i$ that maximises the weighted cluster density $\gweight g(\data\mid\gmean, \gcov)$.
The GMM can be extended to handle data points with Gaussian uncertainty in the following way.
For the data point $x$ with Gaussian error $\sigma$, the probability density function is
\begin{equation}
p(x, \sigma \mid \lambda) = \sum w_i g(x \mid \mu_i, (\gcov + \sigma)).
\end{equation}

\begin{figure*}
    \centering
    \includegraphics[width=\textwidth]{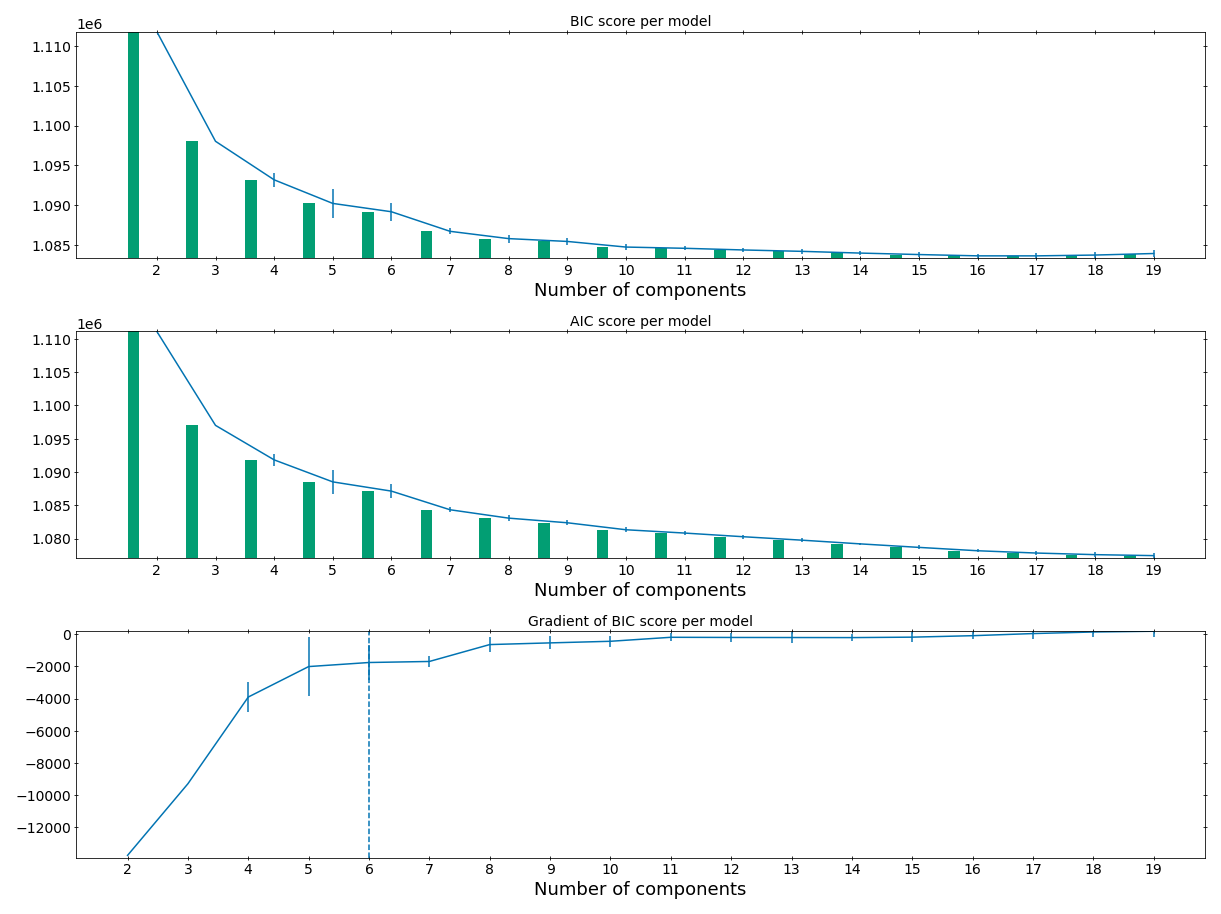}
    \caption{%
    \refereenew{To determine the most appropriate number of components ($\nc$) in the clustering, a standard technique is used that considers the 
    \emph{Bayesian Information Criterion}
    \citep[BIC,][upper panel]{schwartz1978}
    and the \emph{Akaike Information Criterion} \citep[AIC,][middle]{akaike1974}
    as the number of components is varied.
    Specifically, the \emph{Elbow method} \citep{thorndike1953}
    is used in which the information criteria
    are weighed against the model complexity.
    See text for further details.
    In the bottom panel, the gradient of the upper panel is shown and the vertical dashed line
    marks the number of components determined with
    the Kneedle algorithm \citep{kneedle},
    resulting in $\nc = 6$.%
    }}%
    \label{fig:bicgradient}
\end{figure*}

We choose to look for clustering tendencies in the following $D=6$ dimensions: the chemical dimensions of the iron and $\alpha$-element abundances \feh and \alphafe, the kinematic dimensions of the three orbital actions \Jr, \Jphi, and \Jz, and the stellar age $\tau$. One advantage of using Gaussian mixture models over other clustering algorithms is that it is insensitive to different scales used among the dimensions, allowing us to mix dimensions of different units without a dependence on the scaling or choice of normalisation.
We performed a Hopkins cluster tendency test \citep{hopkins1954} as an initial null hypothesis test to assess if any clustering is present in this dimensional subspace and we could reject the null hypothesis that there were no meaningful clusters.

For the GMM, we use the implementation provided in the \textsc{Python} \texttt{pyGMMis} library \citep{pygmmis}, which uses the technique known as Extreme Deconvolution by \citet{bovy2011} to handle data points with uncertainty. To initialise the model,
we use the implementation provided in the \textsc{Python} \texttt{scikit-learn} \citep{pedregosa2011} module called \texttt{GaussianMixture}, which does not support data with uncertainty, but is used since it robustly finds a good initial clustering for \texttt{pyGMMis}. For both GMM implementations, the parameters of the Gaussian mixture models are obtained using the Expectation-Maximization algorithm \citep{dempster1977}.
\begin{figure*}
	\centering
	\includegraphics[width=\textwidth]{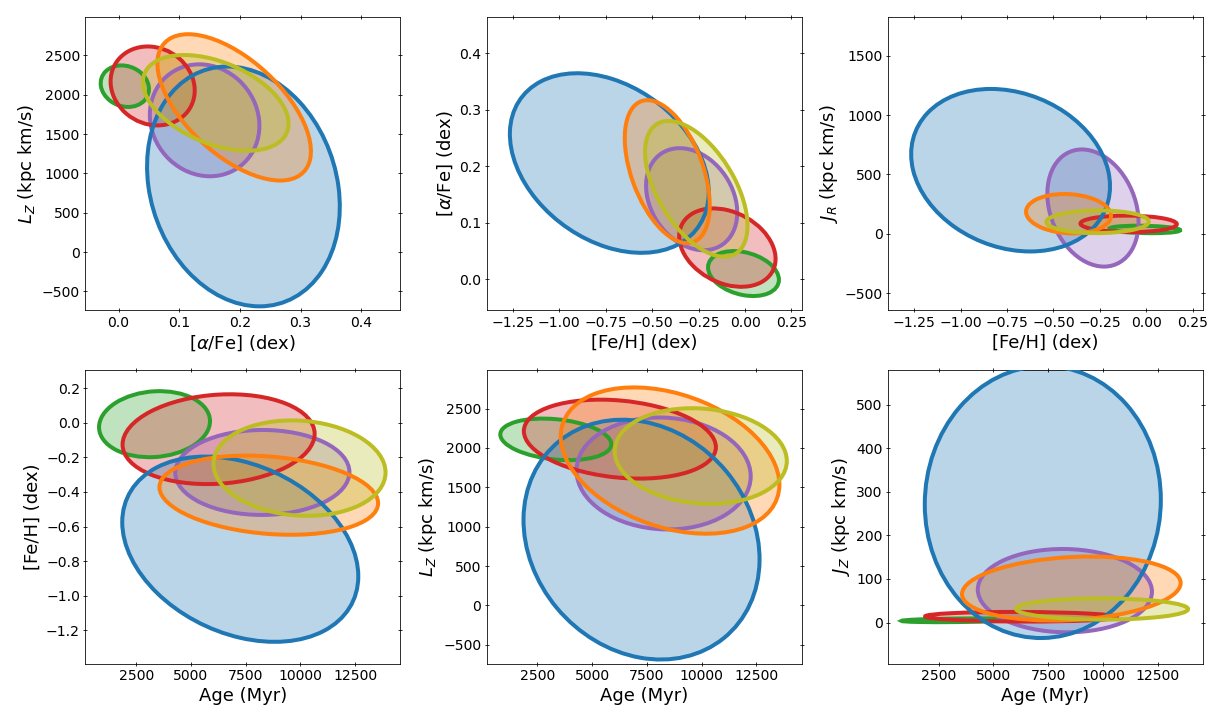}
	\caption{Ellipses showing the approximate regions of each component found in the sample, see Table~\ref{tab:bestmodel}. 
	The principal axes of each ellipse correspond to the eigenvectors of the submatrix of the covariance matrix \gcov corresponding to the two dimension of the given plane as noted by the axis labels. The resulting ellipse is scaled such that its area represents a confidence level of $\SI{95}{\percentage{}}$ or 2 standard deviations.}
	\label{fig:cartoon}
\end{figure*}

\begin{figure*}
	\centering
	\includegraphics[width=\textwidth]{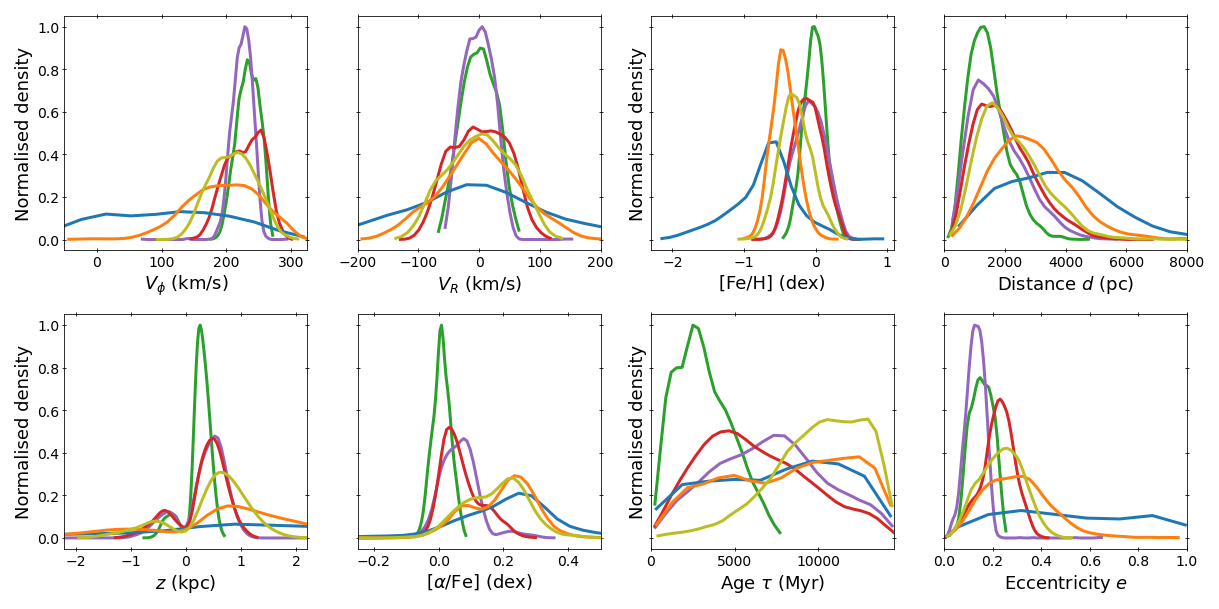}
	\caption{Histograms of different properties for different components, see Table~\ref{tab:bestmodel}.}
	\label{fig:componenthistogram}
\end{figure*}

A free hyperparameter for GMM is the number of clusters \nc. We select the number of clusters \nc by computing the clustering for $\nc=[2$--$20]$ clusters and then doing a comparative evaluation using the Bayesian Information Criterion
\citep[BIC,][]{schwartz1978}, \refereenew{see \figref{fig:bicgradient}}.
The BIC of a model is computed by subtracting a model complexity term from the log-likelihood, thus using the log-likelihood to evaluate the given model while penalising the complexity of the model. The BIC score is fairly smooth and monotone in our case for increasing values of \nc, but sudden changes can be seen in the slope of the BIC score for certain values of \nc. 
In order to select the number of clusters \nc based on the computed BIC scores, we use the so-called
\refereenew{\emph{Elbow method}} \citep{thorndike1953}
in order to choose the number of clusters $\nc$ where the diminishing returns in log-likelihood are no longer worth the additional cost of adding another cluster. We run $10$ iterations of the clustering analysis to assess the numerical variability in the BIC score for different \nc.
Lastly, in order to computationally \refereenew{find the best number of clusters} in the Elbow method,
we use the Kneedle algorithm \citep{kneedle} to locate where a shift in slope in the gradient of the BIC occurs.
This results in a choice of $\nc=6$.

To evaluate the robustness of the classification for the individual stars, we run the clustering assignment $25$ times and look at the variation in the labels.
We compute the confusion matrix pairwise between different labelling for different runs.
Each confusion matrix is an $\nc \times \nc$ matrix, where entry $(i, j)$ is the number of stars with label $i$ in the first clustering and label $j$ in the second clustering. If the two clustering are mostly the same, up to a renumbering of the clusters, then this confusion matrix will contain $\nc$ large and $\nc^2-\nc$ smaller entries.
This is useful because the ordering of the labels is not guaranteed to be identical between different runs, as the algorithm can find the same clustering but assign them different numeric labels in different runs.
We therefore use the Hungarian algorithm/Kuhn-Munkres algorithm \citep{kuhn1955,munkres1957} to find the best matching between the two labels, so we can rename one to match the other.
The off-diagonal elements in the confusion matrix are all no more than a quarter of the value of every component and we thus conclude that we robustly find the same six components in the sample. 
For each target, we will in the following use its majority vote or most frequent label across the different runs.

            \begin{table*}
            \centering
            \caption{Weighted best-fit Gaussian mixture model for the sample. The ordering of the dimensions are \{\feh, \alphafe, \Jr, \Lz, \Jz, $\tau$\} with the following units \{dex, dex, \si{\kilo pc \, \kilo\metre\per\second}, \si{\kilo pc \, \kilo\metre\per\second}, \si{\kilo pc \, \kilo\metre\per\second}, Myr\}.}
            \label{tab:bestmodel}
            \begin{tabular}{l l l l c}
            \hline
            \hline
             Component & Colour & Weight $\gweight$ & Means $\gmean$ & Covariances $\gcov$  \\
            \hline
            
  &  Green  &  $0.202$  &  $\begin{pmatrix*}[r]
  -0.01\\
  0.01\\
  36.92\\
  2109.80\\
  5.27\\
  3337.33\\
\end{pmatrix*}$  &  $\begin{pmatrix*}[r]
  0.018 & -0.001 & -0.604 & -6.800 & -0.070 & 22.098\\
  -0.001 & 0.001 & 0.004 & -0.727 & 0.029 & 27.260\\
  -0.604 & 0.004 & 497.758 & 1505.476 & 4.236 & -3133.780\\
  -6.800 & -0.727 & 1505.476 & 34880.650 & -33.504 & -78820.045\\
  -0.070 & 0.029 & 4.236 & -33.504 & 9.180 & 1690.725\\
  22.098 & 27.260 & -3133.780 & -78820.045 & 1690.725 & 3210358.170\\
\end{pmatrix*}$  \\
  &  Purple  &  $0.140$  &  $\begin{pmatrix*}[r]
  -0.29\\
  0.14\\
  217.26\\
  1676.09\\
  73.28\\
  8261.37\\
\end{pmatrix*}$  &  $\begin{pmatrix*}[r]
  0.030 & -0.003 & -14.694 & 8.213 & -1.618 & 3.205\\
  -0.003 & 0.004 & -3.320 & -3.120 & -0.204 & 44.717\\
  -14.694 & -3.320 & 121551.273 & -50902.382 & 2936.030 & -111159.609\\
  8.213 & -3.120 & -50902.382 & 253129.764 & -7358.521 & -71798.966\\
  -1.618 & -0.204 & 2936.030 & -7358.521 & 4562.006 & -4859.783\\
  3.205 & 44.717 & -111159.609 & -71798.966 & -4859.783 & 7908526.066\\
\end{pmatrix*}$  \\
  &  Red  &  $0.318$  &  $\begin{pmatrix*}[r]
  -0.10\\
  0.06\\
  83.91\\
  2112.79\\
  13.43\\
  6263.93\\
\end{pmatrix*}$  &  $\begin{pmatrix*}[r]
  0.034 & -0.003 & -0.348 & -27.794 & -0.275 & 67.991\\
  -0.003 & 0.002 & -0.277 & -2.008 & 0.046 & 43.236\\
  -0.348 & -0.277 & 2408.133 & 2962.321 & -59.983 & -18299.392\\
  -27.794 & -2.008 & 2962.321 & 125862.886 & -79.994 & -224418.673\\
  -0.275 & 0.046 & -59.983 & -79.994 & 60.670 & -1830.532\\
  67.991 & 43.236 & -18299.392 & -224418.673 & -1830.532 & 9616349.444\\
\end{pmatrix*}$  \\
  &  Orange  &  $0.129$  &  $\begin{pmatrix*}[r]
  -0.42\\
  0.19\\
  169.38\\
  1840.36\\
  78.11\\
  8555.41\\
\end{pmatrix*}$  &  $\begin{pmatrix*}[r]
  0.026 & -0.007 & -1.886 & 23.999 & -2.412 & -115.490\\
  -0.007 & 0.008 & 0.417 & -35.244 & 1.296 & 116.889\\
  -1.886 & 0.417 & 13738.824 & -6514.834 & -2239.821 & -6438.707\\
  23.999 & -35.244 & -6514.834 & 432403.704 & -7537.615 & -766646.789\\
  -2.412 & 1.296 & -2239.821 & -7537.615 & 2694.181 & 32014.772\\
  -115.490 & 116.889 & -6438.707 & -766646.789 & 32014.772 & 12464647.318\\
\end{pmatrix*}$  \\
  &  Olive  &  $0.188$  &  $\begin{pmatrix*}[r]
  -0.26\\
  0.16\\
  101.49\\
  1897.31\\
  31.28\\
  9958.14\\
\end{pmatrix*}$  &  $\begin{pmatrix*}[r]
  0.038 & -0.009 & 0.238 & 16.508 & -0.613 & -49.292\\
  -0.009 & 0.007 & -0.005 & -15.797 & 0.074 & 59.818\\
  0.238 & -0.005 & 4557.381 & -2844.794 & -452.335 & -18084.276\\
  16.508 & -15.797 & -2844.794 & 185163.815 & 66.038 & -122173.911\\
  -0.613 & 0.074 & -452.335 & 66.038 & 295.251 & -647.814\\
  -49.292 & 59.818 & -18084.276 & -122173.911 & -647.814 & 7722049.671\\
\end{pmatrix*}$  \\
  &  Blue  &  $0.023$  &  $\begin{pmatrix*}[r]
  -0.73\\
  0.21\\
  535.58\\
  835.41\\
  276.47\\
  7253.11\\
\end{pmatrix*}$  &  $\begin{pmatrix*}[r]
  0.143 & -0.014 & -36.203 & 201.296 & -17.268 & -421.134\\
  -0.014 & 0.013 & -5.346 & -20.182 & 0.200 & 49.699\\
  -36.203 & -5.346 & 233875.431 & -144285.777 & 8323.666 & 112280.321\\
  201.296 & -20.182 & -144285.777 & 1161630.815 & -50611.197 & -671321.920\\
  -17.268 & 0.200 & 8323.666 & -50611.197 & 48621.357 & 14218.293\\
  -421.134 & 49.699 & 112280.321 & -671321.920 & 14218.293 & 14525739.917\\
\end{pmatrix*}$  \\

                \hline
                \end{tabular}
                \end{table*}

% ~~~~~~~~~~~~~~~~~~~~~~~~~~~~~~~~~~~~~~~~~~~~~~~~~~~~~~~~~~~~~~~~~~~~~~~~~~~~~
% BEST-FIT
% ~~~~~~~~~~~~~~~~~~~~~~~~~~~~~~~~~~~~~~~~~~~~~~~~~~~~~~~~~~~~~~~~~~~~~~~~~~~~~
\subsection{The six best-fit components}
The weighted best-fit GMM of the different runs is summarised in Table~\ref{tab:bestmodel} and an overview plot of the covariance matrix can be seen in \figref{fig:cartoon}.
Histograms of various stellar properties for each component can be seen in \figref{fig:componenthistogram}.

The \youngthingreen and \thinpurple components resemble the traditional thin disc of the Milky Way. 
They are similar in the kinematics spaces as both 
are kinematically cold, meaning that the rotation clearly dominates over the stellar velocity dispersion. Their stars
have
near-solar velocities and move in low-eccentricity orbits. Both components have similar metallicities, however, the \thinpurple component extends further up in \alphafe than the \youngthingreen component.
The key difference between the two is found in stellar age where we see that the \youngthingreen component consists of young stars with a mean age of slightly above $3$~Gyr, while the \thinpurple component has predominately more old stars, with a peak in age around $8$~Gyr.
Spatially, we also see the difference that the stars in the \thinpurple component currently are at greater heights away from the Galactic mid-plane compared to stars belonging to the \youngthingreen component.

At the other extreme in age, we find the \haloblue component. The \haloblue component resembles the traditional stellar halo component as it is metal-poor and has above solar-levels of \alphafe.
The \haloblue component contains stars with more eccentric orbits than the other components and it has a greater dispersion in the different kinematic spaces compared the other clusters. The \haloblue component is predominately old stars and it extends to the largest distances of all of the clusters found.

The distributions of the \thickorange and \oldthickolive components seem to be similar to the traditional thick disc component. 
These two components both contain stars with energetic orbits and have relatively broad velocity dispersions. Both extend further up in eccentricity than for example the \thinpurple and \youngthingreen components, however, the distribution of the \thickorange component is broader and more halo-like than that of the \oldthickolive.
Both the \thickorange and \oldthickolive components are both metal-poor and have above solar levels of \alphafe. In age we see the clearest separation of the two: the \oldthickolive component is predominantly old, while the \thickorange component contain stars over the entire age range.

The \youngthickred component seem similar to the \youngthingreen component in action spaces (see \figref{fig:cartoon}) and chemically it resembles the thin disc-like \youngthingreen and \thinpurple components. The orbital eccentricities peak at greater values than that of the \youngthingreen and \thinpurple components, but it is found closer to the Galactic mid-plane than the \thickorange and the \haloblue components.

% ~~~~~~~~~~~~~~~~~~~~~~~~~~~~~~~~~~~~~~~~~~~~~~~~~~~~~~~~~~~~~~~~~~~~~~~~~~~~~
% ASSEMBLY HISTORY
% ~~~~~~~~~~~~~~~~~~~~~~~~~~~~~~~~~~~~~~~~~~~~~~~~~~~~~~~~~~~~~~~~~~~~~~~~~~~~~
\subsection{Assembly history of the components}
With the observed distributions in mind, we interpret the results as follows.

The \youngthingreen component contains the young stars born in the Milky Way recently, formed from the colder gas in the disc which gets gradually more metal-rich over time as regulated by feedback from star formation. They have not yet had to time to have undergone dramatic dynamical heating \citep[e.g.][]{ting2019}, so they are on circular orbits close to the Galactic mid-plane.
The \thinpurple component is in many ways similar to the \youngthingreen component.
It is possible that the \thinpurple component correspond to the old tail of the traditional thin disc, formed early after an inflow of gas \citep[see e.g.][]{chiappini1997, spitoni2019} before the chemical signature and alpha-enhancement of the interstellar medium reached the solar levels.

The velocities and the broad spatial distributions of the \haloblue component are consistent with it being a hot spheroid of metal-poor and primarily old stars. Some targets are counter-rotating and are likely accreted rather than formed in situ.

The \oldthickolive and \thickorange components compose the traditional thick disc.
This is clear from their characteristics in chemistry and kinematics, where there also is some similarities to the \haloblue component.

When working with GMM one has to be wary about interpreting all components as physically different entities. The GMM fits a multitude of Gaussians to the sample distributions, so two physical components may be labelled as three Gaussian components if the overlap in the tails of the two components can be fitted nicely as a separate third Gaussian. Another possibility is that the distribution is very asymmetric and gets divided into a peak and tail Gaussian component.
The finite range in age could cause the asymmetric distribution of a single thick disc component to be fitted as two Gaussians and we think this is the case for the \oldthickolive and \thickorange components.
We computed a GMM with $n_c=5$ to check and here we reidentify most components with approximately the same overall characteristics. The \thickorange and \oldthickolive components in this case are more-or-less merged, supporting this view that these two components represent the same physical stellar population.

The \youngthickred component could correspond to kinematically heated thin disc stars.
These stars have been heated by interactions with satellites or even full merger events.
It is particular peculiar as its age distribution peaks near $\sim5$~Gyr, meaning that cause for the main perturbation is in the more recent Galactic history. This almost coincides with the first pericentre passage of the Sagittaurius dwarf satellite $5.7$~Gyr ago \citep{ruizlara2020}, which could be the impacting factor behind the kinematic heating of these stars.

% ~~~~~~~~~~~~~~~~~~~~~~~~~~~~~~~~~~~~~~~~~~~~~~~~~~~~~~~~~~~~~~~~~~~~~~~~~~~~~
% COMPARE
% ~~~~~~~~~~~~~~~~~~~~~~~~~~~~~~~~~~~~~~~~~~~~~~~~~~~~~~~~~~~~~~~~~~~~~~~~~~~~~
\citet{nikakhtar2021} recently constructed a GMM from the Gaia DR2 $\times$ APOGEE DR16 cross-match using $D=4$ with the dimensions being the iron abundance \feh, and the three measured velocities in Galactocentric Cartesian coordinates
$V_X$, $V_Y$, and $V_Z$. 
Using their GMM they find five components: like for our GMM a traditional thin disc and halo component are unambiguously identified, while they find that the traditional thick disc component is split into three subdivisions. The difference between the three thick disc components show a difference in the distributions of \alphafe and stellar age.
\citet{nikakhtar2021} also produce a set of synthetic mock-catalogues using the FIRE-2 cosmological simulations. By comparing their data-driven cross-match with the synthetic cross-matches, they conclude that one of the thick disc components represents thin disc stars heated by interactions with Galactic satellites, similar to our \youngthickred component, while the remaining two components represent the velocity asymmetry of the alpha-enhanced thick disc.
Action is not as sensitive to the velocity asymmetry as velocity spaces, but we anyhow find a different splitting of the traditional thick disc component due to our choice of using age as a dimension as described above.

% ~~~~~~~~~~~~~~~~~~~~~~~~~~~~~~~~~~~~~~~~~~~~~~~~~~~~~~~~~~~~~~~~~~~~~~~~~~~~~
% MERGER REMNANTS
% ~~~~~~~~~~~~~~~~~~~~~~~~~~~~~~~~~~~~~~~~~~~~~~~~~~~~~~~~~~~~~~~~~~~~~~~~~~~~~
\section{Age of merger remnants}
Within the standard picture of cosmology the bulk of the stellar halo of a galaxy like the Milky Way is predicted to be dominated by a small number of massive early-accreted dwarf galaxies \citep[e.g.][]{bullockjohnston2005, deluciahelmi2009}.
One of the major discoveries \january{made possible after} the second data release from the Gaia space observatory was that a significant fraction of the halo stars in the solar neighbourhood seem to belong to the same substructure, known in the literature as the debris from a system named Gaia-Enceladus-Sausage (GES) or segments thereof \citep{helmi2018, belokurov2018, myeong2019} which merged with the nascent Milky Way. 
The stars born in the GES system have \refereenew{highly eccentric orbits and appear to have lower \alphafe values for fixed metallicity than the in-situ populations}.
As this merger had a mass ratio of $\sim4$:$1$ \citep{helmi2018, gallert2019}, it is believed that it was the last major merger event that the Milky Way has undergone and that the merger with GES triggered star formation in our early Galaxy.
Stars originating from the GES should be as old as the merger or older, so the age distribution of these stars work as a proxy for when the merging event took place and allow us to better understand the shaping processes in the early Milky Way.

An upper limit on when the major merger event occurred can be set by the study of the bright asteroseismic halo target $\nu$~Indi 
\citep{chaplin2020}. This star was born in the early Milky Way prior to the event and its orbit was heated by the gravitational interactions of the merger. The age of $\nu$~Indi was computed to be $11.0 \pm 1.5$~ Gyr using the asteroseismic constraints from the individual oscillation modes computed from short-cadence TESS time series. 
Using lightcurves from \kepler or TESS along with spectra from the APOGEE and/or LAMOST surveys, asteroseismic constraints seem to point in the direction of the merger happening at $\sim8$~Gyr ago \citep{montalban2021,matsuno2021,grunblatt2021,borre2021}, with the largest sample being that of 8 \kepler and \ktwo halo stars identified as remnants of GES used in the study of \citet{borre2021} yielding a population age of $9.5^{+1.2}_{-1.3}$~Gyr.

% Our summary results
Using our GMM, we could not recover a specific component belonging to that of the GES or other halo substructure. This is most likely a selection bias due to the low density of stars belonging to the Galactic halo in our sample.
\refereenew{Other studies have performed cuts in kinematic space to remove stars with thin-disc and thick-disc kinematics and using similar methods as in this study to successfully find a GES component \citep[][]{liang2021,buder2022}.}
\citet{helmi2018} defines their selection of GES stars in kinematic space as $\SI{-1500}{\km\squared\per\second\squared} < L_Z < \SI{150}{\km\squared\per\second\squared}$ along with requiring an orbital binding energy $E > -1.8 \times 10^5\si{\km\squared\per\second\squared}$.
Choosing the targets that fulfil %or within their uncertainties fall within 
this selection criteria leaves us with a subsample of 384 stars with a median stellar age of $\tau_{\textup{GES, q50}}=9.3$~Gyr with the 16th and 84th quantile being $(\tau_{\textup{GES, q16}}, \tau_{\textup{GES, q84}}) = (4.7, 12.1)$~Gyr.
We note that this selection allows the inclusion of younger, seemingly disk stars as there are 98 targets with metallicities $\feh > -0.5$, the effect of which moves the population age distribution towards slightly younger ages.

\citet{myeong2019} identified GES targets using a criteria in action space and eccentricity defined as $L_Z / J_{\textup{total}} < 0.07$, $(J_R - J_Z) / J_{\textup{total}} < -0.3$ and $e \sim 0.9$. Using this action selection we find a result similar to the above from 79 stars with a median stellar age of $\tau_{\textup{GES,q50}}=9.9$~Gyr with $(\tau_{\textup{GES,q16}}, \tau_{\textup{GES,q84}}) = (6.0, 12.3)$~Gyr. 
In contrast to these definitions in kinematic space, \citet{montalban2021} use a chemical criteria of $[\textup{Mg/Fe}] < -0.2 \feh + 0.05$ along with an orbital eccentricity requirement of $e > 0.7$.
This selects 83 stars in our sample with $\tau_{\textup{GES,q50}}=9.6$~Gyr with the 16th and 84th quantile being $(\tau_{\textup{GES,q16}}, \tau_{\textup{GES,q84}}) = (5.7, 12.5)$~Gyr.
Overall, these different ways of identifying accreted GES stars yield similar population ages of between $9$ and $10$~Gyr, consistent with a merger time between $8-10$~Gyr ago.
This is in agreement with similar studies \citep{matsuno2021, grunblatt2021,borre2021} and gives further credence to the argument that the GES was accreted by the Milky Way less than 10 Gyr ago.

Along side the discovery of GES, multiple other smaller halo substructures have been uncovered. One of these is dubbed Sequoia \citep{myeong2019} and these associated stars are found to have a more retrograde motion on less eccentric orbits and a lower metallicity than GES stars. \citet{myeong2019} identified this component in action and eccentricity space as $L_Z / J_{\textup{total}} < -0.5$, $(J_R - J_Z) / J_{\textup{total}} < 0.1$ and $e \sim 0.6$. Using this criteria and quantifying the eccentricity criteria as $0.45 < e < 0.75$, we find 16 stars with a $\tau_{\textup{Sequoia,q50}}=8.9$~Gyr with $(\tau_{\textup{Sequoia,q16}}, \tau_{\textup{Sequoia,q84}}) = (7.5, 11.4)$~Gyr. 
One suggested origin for this substructure is that Sequoia was a smaller satellite to GES. However, if the GES galaxy had a radial metallicity gradient, which could be expected if the system was of comparable size to that of the Large Magellanic Cloud, then the outskirts of the galaxy should be more metal poor than its central parts and the Sequoia stars could origin from this part of the GES galaxy \citep{koppelman2019}. The outer material of GES would be accreted during the early mass transferring interactions and could be on more retrograde orbits, as is observed for the Sequoia stars.
The age of this component is similar but a bit younger than the age of the GES component, and the age distribution points towards a time of the accretion of $7.6$~Gyr ago. This small age difference can be explained in both origin scenarios e.g. by inside-out formation in the GES system, so these ages alone cannot be used to distinguish between them and need to be combined with more careful modelling of the merger event. 

Another consequence of a major merger event is the perturbation it caused to the orbits of the existing stars in the early Milky Way.
\citet{dimatteo2019} and \citet{amarante2020tail} found stars on heated orbits that are too metal-rich to be considered part of the accreted halo.
This is supported by the study of \citet{belokurov2020}, which also reported the finding of a populations of metal-rich stars on highly eccentric orbits, which they dubbed the Splash stars. They found similar Splash structures in hydrodynamical simulations where the host galaxy underwent a major merger, and because this component connects smoothly to the traditional thick disc in different properties, they conclude that these stars belong to the proto-galactic disc of the Milky Way, which got heated during the major merger event of GES.

\citet{belokurov2020} found that the cleanest selection of the Splash stars are $-0.7 < \feh < -0.2$ and $v_{\phi} < 0$. Using this simple selection, we have $45$ stars in our sample with $\tau_{\textup{Splash,q50}}=10.3$~Gyr with $(\tau_{\textup{Splash,q16}}, \tau_{\textup{Splash,q84}}) = (6.0, 12.3)$~Gyr. All but 7 of these stars belong to our \youngthickred component.
In contrast to the findings of \citet{belokurov2020}, we find that these Splash stars are slightly older than stars accreted from GES, but similarly we find that the age distributions of both populations are comparable.

The Splash stars are believed to have been born in the Milky Way prior to a massive ancient accretion event which drastically altered their orbits. However, \citet{amarante2020} find from hydrodynamical simulations
that the Splash stars can also be reproduced in the absent of a major merger if the star formation in the early galaxy is driven by what they refer to as clumpy formation. In this case, the supernova explosions inject thermal energy to the interstellar medium
following the blastwave implementation of \citet{stinson2006}.
These two scenarios are not mutually exclusive and our study opens a path for future studies of these open formation scenarios using precise asteroseismic ages.

% ~~~~~~~~~~~~~~~~~~~~~~~~~~~~~~~~~~~~~~~~~~~~~~~~~~~~~~~~~~~~~~~~~~~~~~~~~~~~~
% SUMMARY
% ~~~~~~~~~~~~~~~~~~~~~~~~~~~~~~~~~~~~~~~~~~~~~~~~~~~~~~~~~~~~~~~~~~~~~~~~~~~~~
\section{Summary}
\label{sec:conclusion}
In this study we determine precise physical, chemical, and kinematic properties stellar properties for the largest asteroseismic sample of evolved stars with determined stellar ages to date containing \noofstars stars.
These stars all have photometry from 2MASS, available spectroscopy from either APOGEE DR16, GALAH DR3, or LAMOST DR5, astrometry from \gaia EDR3, and the global asteroseismic parameters from either \kepler, K2, or \corot. The stellar properties were inferred using these observables in a Bayesian scheme using \basta.
This data set is available in online and described in Table~\ref{tab:cat}.

Using this sample, we study overall trends as a function of spatial position in the Galaxy.
We find a declining radial age gradient as we move outward in the Galactic disc with a linear slope of about $-0.80$~Gyr~kpc$^{-1}$ for stars close to the Galactic mid-plane $|z|<1$~kpc. We find the approximately the same declining trend in age for different heights away from the mid-plane.
Interestingly, we find an asymmetric trend in stellar age in the outer region of the Galaxy ($\galr > 10$) when we separate the sample into stars below and above the Galactic mid-plane, with stars above the plane being older than stars below.
For the vertical age gradient, we see that at $7 < \galr < 10$~kpc a linear slope is about $0.21$~kpc~Gyr$^{-1}$ or inversely $4.8$~Gyr~kpc$^{-1}$ for stars close to the mid-plane, in agreement with the study of \citet{casagrande2016} for similar stars in the solar cylinder. For the inner region of the Galaxy, this age gradient is close to non-existing, while our results for the outer region of the Galaxy is similar to the result of the solar vicinity with a slope of $0.32$~kpc~Gyr$^{-1}$ or inversely $3.1$~Gyr~kpc$^{-1}$. However, the result for the outer regions is affected by the hemisphere age asymmetry we noticed when studying the radial age gradient.
Above a height of $|z|=1$~kpc, we see a saturation in age at all radii, which sets in at a lower height for the inner region (~$|z|\sim1$~kpc) than for the solar neighbourhood (~$|z|\sim1.4$~kpc) or the outer regions (~$|z|\sim1.5$~kpc). This saturation is likely caused by the flaring of the younger stellar populations, in agreement with inside-out formation of the Milky Way.

In order to study the stellar substructure in the Galactic disc, we use Gaussian mixture models as a data-driven clustering approach for the stars in our sample.
We derive a best-fit, 6-component Gaussian Mixture model of the sample using the following observable subset: the chemical dimensions of \alphafe and \feh, the kinematic dimensions of the orbital actions \Jr, \Lz, and \Jz, and stellar age $\tau$.
These components and their details is presented in Table~\ref{tab:bestmodel} and our interpretation can be summarised as follows: 
\begin{itemize}
    \item The \haloblue component contains stars from the Milky Way stellar halo. These stars are kinematically hot, extend out to greater distances than stars belonging to other components and they are metal-poor and alpha-enhanced. The stars are believed to be remnants from earlier satellite galaxies, which have merged with the early Milky Way as stars originating from the Gaia-Enceladus-Sausage or Sequoia galaxy are members of the component.
    \item The \oldthickolive and \thickorange components contain the stars from the traditional Milky Way thick disc. These stars show great dispersions in velocities and orbit eccentricities and they are metal-poor and alpha-enhanced. The cause of the splitting of this physical galactic component into two components by the GMM is believed to be because the thick disc shows an asymmetric distribution in age, which in this case is then better fitted as two Gaussians: one around the peak (\oldthickolive) and one for the more populous tail (\thickorange).
    \item The \youngthickred component correspond to the kinematically-heated thin disc. These stars look like the traditional thin disc in chemistry space as they have close to solar-levels in metallicity and \alphafe. However, their velocity and action distributions are broader than the thin-disc-like \youngthingreen and \thinpurple components. These stars have most likely been heated by gravitational interactions with satellite galaxies. Because the peak in age for this component is around $5$~Gyr, likely perturbation candidates are the Sagittarius or Gaia-Enceladus-Sausage dwarf galaxies.  
    \item The \thinpurple and \youngthingreen components contains traditional thin disc stars. These stars have alpha-to-iron ratios close to solar levels and orbit the Galactic centre similarly to the Sun in near-circular orbits close to the Galactic mid-plane. These stars are consistent with recent star formation in the cold gas near to the Galactic mid-plane.
\end{itemize}
The resulting model can be used to either classify and study the stars within the data set or to identify members of the components in any data set where the subset of dimensions are available.

\section*{Acknowledgements}
% Thank people
\refereenew{The authors thank Dr.\@ Fiorenzo Vincenzo and the anonymous referee for their careful reading and feedback of the manuscript, which has led to a better presentation of the results.}
% Co-author funding
A.S. acknowledges support from the European Research Council Consolidator Grant funding scheme (project ASTEROCHRONOMETRY, G.A. n. 772293, \url{http://www.asterochronometry.eu}). 
M.H. acknowledges support from NASA through the NASA Hubble Fellowship grant HST-HF2-51459.001 awarded by the Space Telescope Science Institute, which is operated by the Association of Universities for Research in Astronomy, Incorporated, under NASA contract NAS5-26555.
Funding for the Stellar Astrophysics Centre is provided by The Danish National Research Foundation (Grant agreement No.~DNRF106).

% 2MASS
This publication makes use of data products from the Two Micron All Sky Survey, which is a joint project of the University of Massachusetts and the Infrared Processing and Analysis Center/California Institute of Technology, funded by the National Aeronautics and Space Administration and the National Science Foundation.
% Kepler and K2
This paper includes data collected by the \kepler mission, the \ktwo mission, and the \corot space mission. Funding for the \kepler mission and \ktwo mission are provided by the NASA Science Mission directorate.
% COROT and COROGEE
The CoRoT space mission, launched on December 27 2006, was developed and operated by CNES, with the contribution of Austria, Belgium, Brazil, ESA (RSSD and Science Program), Germany and Spain. 

% APOGEE
The work is based on data acquired through the surveys APOGEE, GALAH, and LAMOST.
Funding for the Sloan Digital Sky Survey IV has been provided by the Alfred P. Sloan Foundation, the U.S. Department of Energy Office of Science, and the Participating Institutions. 
The GALAH survey is based on observations made at the Australian Astronomical Observatory, under programmes A/2013B/13, A/2014A/25, A/2015A/19, A/2017A/18. We acknowledge the traditional owners of the land on which the AAT stands, the Gamilaraay people, and pay our respects to elders past and present. Parts of this research were conducted by the Australian Research Council Centre of Excellence for All Sky Astrophysics in 3 Dimensions (ASTRO 3D), through project number CE170100013.
Guoshoujing Telescope (the Large Sky Area Multi-Object Fiber Spectroscopic Telescope; LAMOST) is a National Major Scientific Project built by the Chinese Academy of Sciences. Funding for the project has been provided by the National Development and Reform Commission. LAMOST is operated and managed by the National Astronomical Observatories, Chinese Academy of Sciences

% Gaia
This work has made use of data from the European Space Agency (ESA) mission {\it Gaia} (\url{https://www.cosmos.esa.int/gaia}), processed by the {\it Gaia} Data Processing and Analysis Consortium (DPAC, \url{https://www.cosmos.esa.int/web/gaia/dpac/consortium}). Funding for the DPAC has been provided by national institutions, in particular the institutions participating in the {\it Gaia} Multilateral Agreement.

% SIMBAD
This research has made use of the SIMBAD database, operated at CDS, Strasbourg, France—2000, A\&AS,143,9. This research has made use of the VizieR catalogue access tool, CDS, Strasbourg, France. The original description of the VizieR service was published in A\&AS 143, 23. 
% grendel
The numerical results presented in this work were obtained at the Centre for Scientific Computing, Aarhus \url{http://phys.au.dk/forskning/cscaa}.

% python, numpy, scipy, matplotlib, galpy, astropy
This research made use of the following software: \texttt{Python~3} \citep{python}, \texttt{numpy} \citep{numpy}, \texttt{scipy} \citep{scipy}, \texttt{matplotlib} \citep{matplotlib}, \texttt{kneed} \citep{kneedle}, \texttt{pyGMMis} \citep{pygmmis}, \texttt{galpy} \citep{galpy}, \texttt{astropy} \citep{astropyi, astropyii}, and \basta \citep{silvaaguirre2015,aguirreborsenkoch2021}.

\section*{Data availability statement}
Jupyter notebooks and related online material can be found at the git repository \url{https://github.com/amaliestokholm/gestalt} and in the online supplementary material.

%%%%%%%%%%%%%%%%%%%%%%%%%%%%%%%%%%%%%%%%%%%%%%%%%%


\bibliographystyle{mnras}
\bibliography{main}

%%%%%%%%%%%%%%%%%%%%%%%%%%%%%%%%%%%%%%%%%%%%%%%%%%

%%%%%%%%%%%%%%%%% APPENDICES %%%%%%%%%%%%%%%%%%%%%

\appendix

\section{Output catalogue}
\label{app:cat}
In Table~\ref{tab:cat}, we detail the columns in the output available online.
We give the column name, description and the unit of the quantity where applicable.

\begin{table*}
\caption{Format of provided catalogue. We give the names, description and unit of each provided quantity.}\label{tab:cat}
\begin{tabular}{lcc}
\hline
{Name} & {Description} & {Unit} \\
\hline
\texttt{TMASS\_ID} & unique ID for the 2MASS catalogue & -- \\
\texttt{KIC\_ID} & unique \kepler ID & -- \\
\texttt{EPIC\_ID} & unique \ktwo ID & -- \\
\texttt{SOURCE\_ID\_GAIA} & unqiue source ID for \gaia eDR3 & -- \\
\texttt{LABEL} & which asteroseismic campaign/mission the star was observed in & -- \\
\texttt{SPECTROSURVEY} & which spectroscopic survey is used & -- \\
\texttt{COMPONENT} & the majority label of the clustering & -- \\
\textbf{Input to \basta}&&\\
\texttt{PHASE} & evolutionary phase & -- \\
\texttt{DNU} & the large frequency separation & $\mu$Hz \\
\texttt{NUMAX} & the frequency of maximum power  & $\mu$Hz \\
\texttt{JMAG\_TMASS} & the $J$ magnitude from 2MASS & mag \\
\texttt{HMAG\_TMASS} & the $H$ magnitude from 2MASS & mag \\
\texttt{KMAG\_TMASS} & the $K_s$ magnitude from 2MASS & mag \\
\texttt{TEFF} & effective temperature & K \\
\texttt{M\_H} & metallicity & dex \\
\texttt{ALPHA\_FE} & \alphafe & dex \\
\texttt{RA\_GAIA} & right ascension from \gaia & deg \\
\texttt{DEC\_GAIA} & declination from \gaia & deg \\
\texttt{PMRA\_GAIA} & proper motion in RA from \gaia & mas/yr \\
\texttt{PMDEC\_GAIA} & proper motion in DEC from \gaia & mas/yr \\
\texttt{PARALLAX\_GAIA} & parallax from \gaia corrected for a zero-point offset & mas \\
\texttt{UNCORRECTED\_PARALLAX\_GAIA} & the uncorrected parallax from \gaia & mas \\
\texttt{ERROR\_UNCORRECTED\_*} & Uncorrected uncertainty for the field & -- \\
\textbf{Flags}&&\\
\texttt{DNU\_PROB\_NEW} & K2GAP flag for \dnu & -- \\
\texttt{QFLG\_TMASS} & photometric flag from \tmass & -- \\
\texttt{CHI2RATIO\_LAMOST} & spectroscopic flag from \lamost  & -- \\
\texttt{TEFF\_FLAG\_LAMOST} & spectroscopic flag from \lamost  & -- \\
\texttt{LOGG\_FLAG\_LAMOST} & spectroscopic flag from \lamost  & -- \\
\texttt{VMIC\_FLAG\_LAMOST} & spectroscopic flag from \lamost  & -- \\
\texttt{MG\_FE\_FLAG\_LAMOST} & spectroscopic flag from \lamost  & -- \\
\texttt{SI\_FE\_FLAG\_LAMOST} & spectroscopic flag from \lamost  & -- \\
\texttt{FE\_H\_FLAG\_LAMOST} & spectroscopic flag from \lamost  & -- \\
\texttt{FLAG\_SINGLESTAR\_LAMOST} & spectroscopic flag from \lamost  & -- \\
\texttt{STARFLAG\_APOGEE} & spectroscopic bitmask flag from \apogee  & -- \\
\texttt{ASPCAPFLAG\_APOGEE} & spectroscopic bitmask flag from \apogee   & -- \\
\texttt{FLAG\_SP\_GALAH} & spectroscopic flag from \galah & -- \\
\texttt{FLAG\_FE\_H\_GALAH} & spectroscopic flag from \galah  & -- \\
\texttt{FLAG\_MG\_FE\_GALAH} & spectroscopic flag from \galah  & -- \\
\texttt{FLAG\_SI\_FE\_GALAH} & spectroscopic flag from \galah  & -- \\
\texttt{ASTROMETRIC\_PARAMS\_SOLVED\_GAIA} & number of astrometric parameters in \gaia  & -- \\
\hline
\end{tabular}
\end{table*}

\begin{table*}
\begin{tabular}{lcc}
\hline
{Name} & {Description} & {Unit} \\
\hline
\textbf{\basta output}&&\\
\texttt{AGE\_BASTA} & stellar age from \basta & Myr \\
\texttt{LPHOT\_BASTA} & stellar luminosity from \basta & solar units \\
\texttt{MASSFIN\_BASTA} & stellar mass from \basta & solar units \\
\texttt{RADIUS\_BASTA} & stellar radius from \basta & solar units\\
\texttt{LOGG\_BASTA} & surface gravity from \basta & solar units \\
\texttt{TEFF\_BASTA} & effective temperature from \basta & solar units \\
\texttt{FE\_H\_BASTA} & \feh from \basta & solar units \\
\texttt{ME\_H\_BASTA} & metallicity from \basta & solar units \\
\texttt{DNUSER\_BASTA} & \dnu from \basta & solar units \\
\texttt{NUMAX\_BASTA} & \numax from \basta & solar units \\
\texttt{DISTANCE\_BASTA} & distance from \basta & solar units \\
\texttt{XINI\_BASTA} & initial hydrogen fraction from \basta & -- \\
\texttt{YINI\_BASTA} & initial helium fraction from \basta & -- \\
\texttt{ZINI\_BASTA} & initial metal fraction from \basta & -- \\
\textbf{Coordinates and dynamics}&&\\
\texttt{XRECT\_GC} & $X$ Cartesian coordinate in Galactocentric frame & kpc \\
\texttt{YRECT\_GC} & $Y$ Cartesian coordinate in Galactocentric frame & kpc \\
\texttt{ZRECT\_GC} & $Z$ Cartesian coordinate in Galactocentric frame & kpc \\
\texttt{U\_HC} & $U$ velocity component in Heliocentric frame & km/s \\
\texttt{V\_HC} & $V$ velocity component in Heliocentric frame & km/s \\
\texttt{W\_HC} & $W$ velocity component in Heliocentric frame & km/s \\
\texttt{R\_GC} & \galr cylindrical coordinate in Galactocentric frame & kpc \\
\texttt{PHI\_GC} & $\phi$ cylindrical coordinate in Galactocentric frame  & -- \\
\texttt{Z\_GC} & $z$ cylindrical coordinate in Galactocentric frame  & kpc \\
\texttt{VR\_GC} & velocity in \texttt{R\_GC} & km/s \\
\texttt{VPHI\_GC} & velocity in \texttt{PHI\_GC} & km/s  \\
\texttt{VZ\_GC} & velocity in \texttt{Z\_GC} & km/s  \\
\texttt{JR} & radial action \Jr & kpc km/s \\
\texttt{LZ} & angular momentum in $z$-direction & kpc km/s \\
\texttt{JZ} & vertical action \Jz & kpc km/s \\
\texttt{THETA\_R} & corresponding angle to \Jr & -- \\
\texttt{THETA\_PHI} & corresponding angle to \Lz & -- \\
\texttt{THETA\_Z} & corresponding angle to \Jz & -- \\
\texttt{ECCENTRICITY} & eccentricity of orbit & -- \\
\texttt{ZMAX} & maximum vertical height of orbit & kpc \\
\texttt{RPERI} & pericentre radius & kpc \\
\texttt{RAP} & apocentre radius & kpc \\
\texttt{ENERGY} & orbital energy & km$^2$ s$^{-2}$ \\
\texttt{ERROR\_*} & Corresponding uncertainty for each field & -- \\
\texttt{UPPER\_ERROR\_*} & Corresponding 84th quantile of the distribution for each field & -- \\
\texttt{LOWER\_ERROR\_*} & Corresponding 16th quantile of the distribution for each field & -- \\
\hline
\end{tabular}
\end{table*}

%%%%%%%%%%%%%%%%%%%%%%%%%%%%%%%%%%%%%%%%%%%%%%%%%%

% Don't change these lines
\bsp	% typesetting comment
\label{lastpage}
\end{document}